\documentclass[reprint, aps, amsmath, amssymb, superscriptaddress, nofootinbib]{revtex4-2}
\usepackage{amsmath}
\usepackage{siunitx}
\usepackage{verbatim}
\usepackage{bm}
\usepackage{ulem}
\usepackage{helvet}
\usepackage[bookmarks, colorlinks=false]{hyperref}
\usepackage{mathtools}
\hypersetup{colorlinks=true, allcolors=blue}
\usepackage{verbatim}
\usepackage{csquotes}
\usepackage{xcolor}
\usepackage{multirow}
\usepackage{subfigure}

\begin{document}

\title{The effect of multi-occupancy traps on the diffusion and retention of multiple hydrogen isotopes in irradiated tungsten and vanadium}

\author{Sanjeet Kaur}
	\email{sanjeet.kaur@ukaea.uk}
	\affiliation{UK Atomic Energy Authority, Culham Campus, Abingdon, Oxfordshire OX14 3DB, United Kingdom}

\author{Daniel R. Mason}
	% \email{daniel.mason@ukaea.uk}
	\affiliation{UK Atomic Energy Authority, Culham Campus, Abingdon,  Oxfordshire OX14 3DB, United Kingdom}

\author{Prashanth Srinivasan}
	% \email{prashanth.srinivasan@ukaea.uk}
	\affiliation{UK Atomic Energy Authority, Culham Campus, Abingdon, Oxfordshire OX14 3DB, United Kingdom}

\author{Stephen Dixon}
	% \email{stephen.dixon@ukaea.uk}
	\affiliation{UK Atomic Energy Authority, Culham Campus, Abingdon, Oxfordshire OX14 3DB, United Kingdom}

\author{Sid Mungale}
	% \email{sid.mungale@ukaea.uk}
	\affiliation{UK Atomic Energy Authority, Culham Campus, Abingdon, Oxfordshire OX14 3DB, United Kingdom}

\author{Teresa Orr}
	% \email{tez.orr@ukaea.uk}
	\affiliation{UK Atomic Energy Authority, Culham Campus, Abingdon, Oxfordshire OX14 3DB, United Kingdom}
    
\author{Mikhail Yu. Lavrentiev}
	% \email{mikhail.lavrentiev@ukaea.uk}
	\affiliation{UK Atomic Energy Authority, Culham Campus, Abingdon, Oxfordshire OX14 3DB, United Kingdom}

\author{Duc Nguyen-Manh}
	% \email{duc.nguyen@ukaea.uk}
	\affiliation{UK Atomic Energy Authority, Culham Campus, Abingdon, Oxfordshire OX14 3DB, United Kingdom}
    \affiliation{Department of Materials, University of Oxford, Oxford, OX1 3PH, UK}


\begin{abstract}
\noindent
We propose a computational scheme for the diffusion and retention of multiple hydrogen isotopes (HI) with multi-occupancy traps parameterized by first principles calculations. We show that it is often acceptable to reduce the complexity of the coupled differential equations for gas evolution by taking the dynamic steady state, a generalisation of the Oriani equilibrium for multiple isotopes and multi-occupancy traps. The \textit{effective} gas diffusivity varies most with mobile fraction when the total gas concentration approximates the trap density. We show HI binding to a monovacancy in vanadium produces a non-monotonic dependence between diffusivity and gas concentration, unlike the tungsten system. We demonstrate the difference between multiple single occupancy traps and multi-occupancy traps in long-term diffusion dynamics. The applicability of the multi-occupancy, multi-isotope model in steady state is assessed by comparison to an isotope exchange experiment between hydrogen and deuterium in self-ion irradiated tungsten. The vacancy distribution is estimated with molecular dynamics, and the retention across sample depth shows good agreement with experiment using no fitting parameters.
\end{abstract}

\maketitle

%%%%%%%%%%%%%%%%%%%%%%%%%%%%%%%%%%%%%%%%%%%%%%%%%%%%%%%%%%%%%%%%%
%%%%%%%%%%%%%%%%%%%%%%%%%%%%%%%%%%%%%%%%%%%%%%%%%%%%%%%%%%%%%%%%%
%%%%%%%%%%%%%%%%%%%%%%%%%%%%%%%%%%%%%%%%%%%%%%%%%%%%%%%%%%%%%%%%%

\section{Introduction} \label{sec:introduction}

\noindent
In any proposed fusion tokamak design, there are many expected processes between the plasma and the surrounding wall to consider \cite{m-rubel}. Hydrogen isotope gas at high temperatures inside a fusion reactor vessel can dissolve into or detach from the plasma-facing materials \cite{m-gilbert, fusion-modelling-physics, hydrogen-in-fusion-and-fission}. Given their relatively small size in a metal lattice, the gas atoms occupy interstitial sites between the host atoms \cite{oriani-hydrogen-in-metals}, and diffuse by hopping between these interstitial sites. As a thermally activated process, the rate of diffusion is given by the Arrhenius relation: migration rate $\propto \text{exp}(-E_m/k_BT)$ for lattice temperature $T$ and a migration activation barrier $E_m$. We are interested in the amount of gas retained by the microstructure, as well as the amount released into the reactor vessel or the coolant subsystem over time. \\

\noindent
Predictions for gas retention in plasma-facing materials will advise the tritium fuel inventory in fusion power plants, as well as the subsequent development of a tritium fuel cycle. Heavy-water reactors are currently the only commercially-viable option for tritium production, which have historically produced no more than three kilograms annually \cite{tritium-production}. Pearson \textit{et al.} \cite{tritium-supply-demand} suggested that the significant uncertainty in future tritium removal facilities, which extract the fuel after production, may cause a limited fuel supply during crucial deuterium-tritium (DT) campaigns, especially for the proposed \textit{International Thermonuclear Experimental Reactor} (ITER) programme. \\

\noindent
The outer fuel cycle (OFC) accounts for breeding and coolant processing as well as trapped inventories in plasma-facing components, such as the first wall and divertor. Therefore, the gas diffusion and retention physics applied to estimate fuel inventory must be sufficiently accurate and applicable beyond the current experimental regime \cite{tritium-fuel-cycle, IFMIF}. Finally, for a self-sustaining fusion reactor, the required tritium breeding ratio (TBR) is determined with reference to all possible inefficiencies in the fuel cycle, including the fraction of T not usefully recovered from reactor components \cite{tritium-fuel-cycle}. \\

\noindent
The accumulation of gas in irradiation defects has been previously observed in ion irradiation and plasma exposure studies \cite{damage_hydrogen_accumulation, nra-bulk-concentration}. The process responsible for accumulation is known as \textit{trapping}, the successful migration of a gas atom into an available trap site. A gas atom may then \textit{detrap} from the trap site by overcoming the defect detrapping energy. Conventionally, the detrapping energies are calculated using first principles methods such as density-functional theory (DFT) \cite{dft-defect-energies}, or from atomistic simulations using an appropriate gas-material interatomic potential. A dynamic steady state is reached when the rates of trapping and detrapping are balanced. Both trapping and detrapping are thermally activated at almost all temperatures, so their rates follow the Arrhenius law and this dynamic steady state is strongly temperature dependent.  \\

\noindent
The standard model for the diffusion and retention of hydrogenic gases was developed by McNabb and Foster \cite{mcnabb1963}. This is the simplest description of the change in mobile gas concentration in space and time due to diffusion and traps of a single type. Each trap is single occupancy, either empty or occupied by one gas atom, and the change in trapped gas concentration over time includes a positive trapping term and negative detrapping term. The model has been successfully implemented in many finite-element codes for hydrogen transport studies, such as the \textit{Finite Element Simulation of Tritium in Materials} ({\tt FESTIM}) \cite{festim} and the \textit{Tritium Migration Analysis Program} ({\tt TMAP}) \cite{tmap8}. \\

\noindent
However, there is a major discrepancy between the single occupancy trap model and DFT calculations, which predict that multiple gas atoms may bind to certain trap types. For example the monovacancy is a significant trap for gases in body-centred cubic (bcc) transition metals. It has been suggested that a monovacancy may store up to six H atoms in these metals once H-H repulsion is considered, possibly more at very low temperatures \cite{heinola-vacancy, dft-vacancies}. The detrapping energy required to eject a single gas atom changes with the number of trapped atoms. Another defect structure, the nanovoid, not only traps multiple H atoms \cite{bcc-nanovoids} but has been shown to stabilise the formation of H$_{2}$ molecules inside the void, forming a hydrogen bubble \cite{nanovoid-bubble}. The retention mechanisms in gas-filled cavities are unlike single occupancy traps \cite{gas-cavities}, so gas retention in complex microstructures must extend beyond the McNabb-Foster model. \\

\noindent
To address this discrepancy, Hodille \textit{et al.} have developed the \textit{Migration of Hydrogen Isotopes in MaterialS} ({\tt MHIMS}-reservoir) code for multi-occupancy traps \cite{mhims}, which treats the binding energy of the $n^{th}$ hydrogen atom to a trap as distinct from the binding energy of the $n+1^{th}$ hydrogen atom to the trap. In this model, the incremental binding energies for one isotope, computed by first principles methods, can be used directly to refine trapped concentration estimates. Schmid \textit{et al.} went further with the {\tt TESSIM-X} code, which models two hydrogen isotopes in multi-occupancy traps \cite{schmid}. Here a computational difficulty arises, as the number of distinct configurations of atoms in a trap increases rapidly with the number of isotopes considered. As a practical solution, {\tt TESSIM-X} makes the approximation that the detrapping energies are the same for each isotope, which reduces the number of trapping/detrapping ordinary differential equations (ODEs). \\

\noindent
The multi-occupancy trap equations may be incorporated into the diffusion equation for mobile gas and solved over time and space, in order to replicate concentration profiles or thermal desorption spectroscopy (TDS) profiles from experiments. For example, cluster dynamics (CD) and kinetic Monte-Carlo (kMC) methods define each gas-defect complex uniquely, such as an empty vacancy, vacancy + 1H, vacancy + 2H etc., and evolve each partial differential equation (PDE) either stochastically in kMC \cite{chen_cd, okmc_Mo}, or numerically in CD \cite{improved_cd}. The transition rate physics employed is the same as in the models previously mentioned. As such, Object-kMC has been shown to agree with {\tt MHIMS} in modelling the thermal desorption of H from W with constant vacancy density \cite{okmc-mhims}. The profiles showed multiple peaks corresponding to sequential detrapping from a single trap type, demonstrating the use of DFT calculations in \textit{predictive} modelling. This work recasts the full multi-occupancy and multi-isotope diffusion and trapping equations in a manner which can be solved in a finite-element solver.
\\

\noindent
In section \ref{sec:theory:multioccupancy}, we first outline the formalism for multi-occupancy trapping with one gas species and show how it reduces to single occupancy trapping, to make a comparison with \cite{mhims,schmid}. We then show how to derive the total trapped concentration and \textit{effective} gas diffusivity under a trapping-detrapping dynamic equilibrium using these equations in section \ref{sec:theory:dynamic_ss}. We demonstrate the applicability of this equilibrium in a range of mobile gas concentrations and temperatures. Finally, we extend the mathematical framework to multiple isotopes in multi-occupancy traps in section \ref{sec:theory:multiisotopes:maths}, and consider the influence of zero-point energy corrections on retention estimates in section \ref{sec:theory:multiisotopes:ZPE}.\\

\noindent
The results sections \ref{sec:results:diff-WvsV} and \ref{sec:results:diff-multivssingle} present the difference between single occupancy and multi-occupancy traps in H effective diffusivity in W and V populated by monovacancies. Sequential D and H gas loading in self-ion irradiated W is simulated with the model and compared to previous experimental work in section \ref{sec:results:experiment}. \\

\noindent
This paper demonstrates a tractable scheme to parameterize and efficiently solve for the diffusion and retention of hydrogen isotopes in simple metals, and demonstrates its utility when integrated into a \textit{Multiphysics Object-Oriented Simulation Environment} ({\tt MOOSE}) \cite{giudicelli2024moose} application with an example calculation over space and time. By decoupling the gas-trap dynamics with trap evolution, we aim to demonstrate the explicit effect of sequential binding on diffusivity and retention, as opposed to a single-occupancy formalism. 

%%%%%%%%%%%%%%%%%%%%%%%%%%%%%%%%%%%%%%%%%%%%%%%%%%%%%%%%%%%%%%%%%
%%%%%%%%%%%%%%%%%%%%%%%%%%%%%%%%%%%%%%%%%%%%%%%%%%%%%%%%%%%%%%%%%
%%%%%%%%%%%%%%%%%%%%%%%%%%%%%%%%%%%%%%%%%%%%%%%%%%%%%%%%%%%%%%%%%

\section{Theory} \label{sec:theory}

%%%%%%%%%%%%%%%%%%%%%%%%%%%%%%%%%%%%%%%%%%%%%%%%%%%%%%%%%%%%%%%%%
%%%%%%%%%%%%%%%%%%%%%%%%%%%%%%%%%%%%%%%%%%%%%%%%%%%%%%%%%%%%%%%%%

\subsection{The multi-occupancy trap}\label{sec:theory:multioccupancy}

\noindent
Atomic gas in a crystal lattice is split into two mutually exclusive populations: \textit{mobile} and \textit{trapped}. The mobile gas concentration, expressed as an atomic fraction, is given by a scalar field $x(\textbf{r}, t)$ for a single gas species, or the vector of scalar fields $\textbf{x}(\textbf{r}, t)$ for multiple gas species. For exposition purposes, we start by only considering a single gas species until section~\ref{sec:theory:multiisotopes}.\\

\noindent
An $n$-occupancy trap may be empty or occupied by up to $n$ gas atoms.  
We identify the occupancy state of a trap containing gas atoms with a state label, $s$. 
The probability that an individual trap at position ${\bf r}$ and time $t$ is in state $s$ is given by $y_s(\textbf{r}, t)$, with $\sum_s y_s(\textbf{r}, t) = 1$. 
The number of gas atoms in state $s$ can be represented by a counting number, $C_s$.
Then, the expected number of gas atoms in the trap is given by $\sum_s C_s y_s$\footnote{Note that for a single isotope, we could have labelled trap states $s$ with the number of gas atoms, ie $s \in \{ 0,1,\ldots,n \}$, in which case $C_s = s$. We keep the formalism of the counting numbers to simplify the multi-isotope case.}.\\

\begin{figure}
    \centering
    \includegraphics[width=\linewidth]{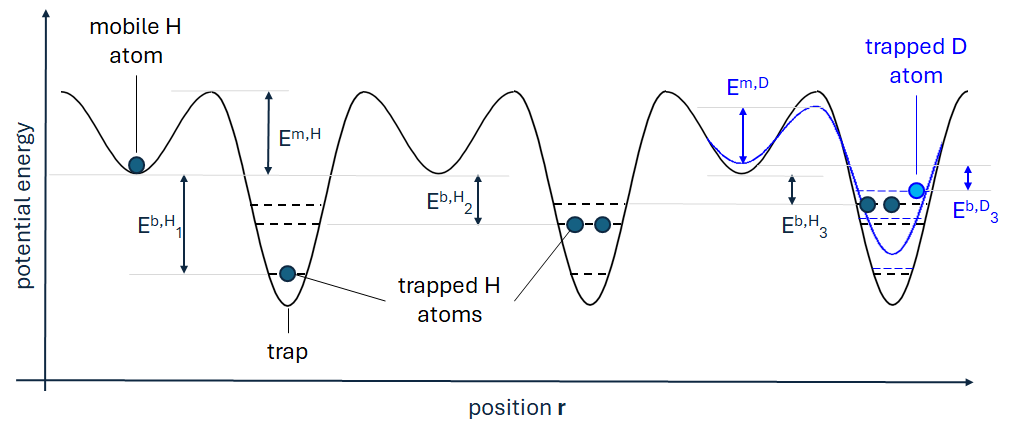}
    \caption{A schematic illustration of a three-level trap. Mobile gas interstitial sites are separated by distance $a$ and migration activation barrier $E_m$. All H atoms in a trapped occupancy state have the same detrapping energy. In this work the detrapping energy from a trap in occupancy state $s$ is the sum of the binding and migration energies. A D atom has a different zero point energy to H, so its migration barrier and binding energy are different.}
    \label{fig:three_level_trap}
\end{figure}

\noindent
The evolution of the probability of a trap being found in state $s$ depends on the rate of trapping into a state with one fewer trapped atoms than $s$, the rate of detrapping from state $s$, and the rate of detrapping from a state with one more trapped atom than $s$,
    \begin{eqnarray*}
        \frac{\partial y_s}{\partial t} &\sim& \mbox{(trapping rate $s_-$)} y_{s_-}  \nonumber\\
                &&  - \mbox{(detrapping rate s)} y_s \nonumber\\
                &&  + \mbox{(detrapping rate $s_+$)} y_{s_+}.
    \end{eqnarray*}    
We define the trapping rate to be proportional to the local number of mobile gas atoms, $x$, and a frequency $k=g D/a^2$, where $g\sim1$ is a trap-specific geometric factor and $D$ is the gas diffusivity given in terms of hop length $a$, migration barrier $E_m$ and attempt frequency $\nu$:
    \begin{equation*}
        D = \frac{ a^2 \nu }{6} \exp \left[-\frac{E_m}{k_BT} \right].
    \end{equation*}
The attempt frequency $\nu$ can be determined by Vineyard's method \cite{vineyard}, using a Nudged Elastic Band calculation~\cite{Henkelman_JCP2000} to find the saddle point, and is often in the order of $10^{13} - 10^{14}$ Hz for light isotopes. Note $\nu \exp \left[-{E_m}/{k_BT} \right]$ is the \textit{total} escape rate from a metastable interstitial site~\cite{Voter_KMC2007}.
The detrapping rate is proportional to the thermal activation rate of an atom leaving the trap. 
In this work we define the detrapping rate for state $s$ to be 
    \begin{equation*}
        p_s = g' \, C_s k \exp \left[-\frac{E^b_s}{k_BT} \right], 
    \end{equation*}
where $g' \sim 6$ is a trap-specific geometric factor and $E^b_s$ is the incremental binding energy of state $s$. Note that $p_s \sim \exp \left[-\frac{E_m+E^b_s}{k_BT} \right]$, consistent with a detrapping energy $E_m+E^b_s$. 
The factor $C_s$ counts the number of gas atoms in the trap, implying that each gas atom is at the same energy level and equally likely to be the next to detrap. Note that this factor is included in Schmid \textit{et al.} \cite{schmid} but omitted in Hodille \textit{et al.} \cite{mhims}. The binding and migration energies are illustrated as potential wells in figure \ref{fig:three_level_trap}. The energy required to detrap from an occupancy state is unique for each isotope due to differences in zero-point energy: this effect is revisited in section \ref{sec:theory:multiisotopes:ZPE}. \\

\noindent
From these considerations, it follows that the time evolution of $\textbf{y}(\textbf{r}, t) = [y_0(\textbf{r}, t),y_1(\textbf{r}, t)\ldots]^T$ is linear in $\textbf{y}$, and can be written as the matrix equation
\begin{equation}
    \label{eqn:time_derivative_y}
    \frac{\partial \textbf{y}}{\partial t} = - {\bf G}[x, T] \textbf{y},
\end{equation}
where ${\bf G}[x, T]$ is a rate matrix which depends on the local mobile gas concentration and temperature, but with no explicit dependence on time.
The trap dynamics are also independent of the trap density. This holds because each trap is treated identically and independently. 
For a single isotope and an $n$-occupancy trap, ${\bf G}[x, T]$ is a simple tri-diagonal square matrix of order $n+1$,

\begin{equation}
\label{eqn:tridiagG}
{\bf G}[x, T] = \left( \begin{array}{ccccc}
                x k     &   -p_1        &               &             &    0        \\
                -x k    &   x k + p_1   &   -p_2      &             &             \\
                        &   -x k        &   x k+ p_2   &   -p_3    &             \\
                        &               &   -x k        &    \ddots   & - p_n      \\
                    0   &               &               &    -x k     &  p_n 
            \end{array}
            \right).
\end{equation}      

\noindent
If the trap density at position ${\bf r}$ and time $t$ is $\rho(\textbf{r},t)$ then the total gas concentration, expressed as an atomic fraction, is
    \begin{equation}
        \label{eqn:totalconc}
        c(\textbf{r},t) = x(\textbf{r},t) + \rho(\textbf{r},t) \, {\bf C} \cdot \textbf{y}(\textbf{r},t).
    \end{equation}

\noindent
where ${\bf C} = [C_0,C_1,\ldots]^T$. Ignoring source terms, boundary conditions and trap evolution, the time evolution of mobile gas and the occupancy probability vector for the trap is given by Fickian diffusion and the exchange between the mobile and trapped populations,
    \begin{eqnarray}
        \label{eqn:multiocc-evolution}
            \frac{\partial x}{\partial t} &=&  \nabla \cdot ( D \nabla  x) + \rho \, {\bf C} \cdot {\bf G} {\bf y}       \nonumber\\                    
            \frac{\partial {\bf y}}{\partial t} &=& - {\bf G} {\bf y}.
    \end{eqnarray}

%%%%%%%%%%%%%%%%%%%%%%%%%%%%%%%%%%%%%%%%%%%%%%%%%%%%%%%%%%%%%%%%%

\subsubsection{Comparison to the single occupancy trap}\label{sec:theory:multioccupancy:single}

\noindent
There has been extensive work on the single occupancy trap and its influence on gas diffusion in metals. McNabb and Foster \cite{mcnabb1963} modelled a single occupancy trap with a distinct trapping rate $k$ and detrapping rate $p$. In their equations, the trapped gas concentration, often labelled $c_t$ in the literature, evolves in time as
\begin{equation}
\label{eqn:mcnabb}
    \begin{gathered}
    \frac{\partial x}{\partial t} =  \nabla \cdot (D \nabla x) - \frac{\partial c_t}{\partial t}
    \\
    \frac{\partial c_t}{\partial t} = k x(\rho-c_t) - p c_t.
    \end{gathered}
\end{equation}

\noindent
By comparing to the equations above, we see that this is indeed the limit of the multi-occupancy trap equations \ref{eqn:multiocc-evolution} with $n=1$, if we identify  
    \begin{eqnarray}
        \label{eqn:Gsingleoccu}
        {\bf C} &=& \left[ 0,1 \right]^T                        \nonumber\\
        c_t &=& \rho \, {\bf C} \cdot {\bf y}                   \nonumber\\
        {\bf G}[x, T] &=& \left( \begin{array}{cc}
            x k & -p \\
            -x k & p
        \end{array} \right ).
    \end{eqnarray}

\noindent
Impurity atoms may be single occupancy traps for H, for example C in a substitutional lattice site in bulk W binds to H with binding energy 1.25eV as calculated using DFT in \cite{impurity-H}.

%%%%%%%%%%%%%%%%%%%%%%%%%%%%%%%%%%%%%%%%%%%%%%%%%%%%%%%%%%%%%%%%%

\subsubsection{Multiple trap types}\label{sec:theory:multioccupancy:multi_trap_types}
\noindent
The extension of equations \ref{eqn:totalconc}-\ref{eqn:Gsingleoccu} to multiple trap types is trivial: we can define a rate matrix ${\bf G}_j$ for each trap $j$, with one trapping rate $k_j$ and a set of detrapping rates $\{p_{s,j}\}$ across occupancy states $s$. Each trap will have its own set of counting numbers ${\bf C}_j$. The total gas concentration is then $c(\textbf{r},t)  = x(\textbf{r},t)  + \sum_j \rho_j(\textbf{r},t)   \, {\bf C}_j  \cdot \textbf{y}_j(\textbf{r},t)$. We return to multiple trap types in section \ref{sec:theory:dynamic_ss:multi-dynamics}.

%%%%%%%%%%%%%%%%%%%%%%%%%%%%%%%%%%%%%%%%%%%%%%%%%%%%%%%%%%%%%%%%%
%%%%%%%%%%%%%%%%%%%%%%%%%%%%%%%%%%%%%%%%%%%%%%%%%%%%%%%%%%%%%%%%%

\subsection{The dynamic steady state and effective diffusivity} \label{sec:theory:dynamic_ss}

\noindent
${\bf G}$ is rank-deficient, because it is necessary for trapping and detrapping rates to balance, $\sum_{s'}G_{s's} = 0$, in order to conserve particle number.
Therefore ${\bf G}$ supports a zero eigenmode. We interpret the zero eigenmode as the dynamic steady state probability vector $\textbf{y}^{\rm eq}(\textbf{r}, t)$, for which 
    \begin{equation*}
        \frac{\partial {\bf y}^{\rm eq}}{\partial t} = - {\bf G} {\bf y}^{\rm eq} = \textbf{0}.
    \end{equation*}
We discuss our approach to accurately calculating ${\bf y}^{\rm eq}$ in the appendix \ref{sec:appendix:A}.
The eigenvalue spectrum of the matrix ${\bf G}[x, T]$ may be bounded with Gershgorin's circle theorem \cite{Gershgorin_mathworld}: the magnitude of the $s^{th}$ eigenvalue, $\lambda_s$, is bounded by $|G_{ss}| \pm \sum_{s' \neq s}|G_{ss'}|$. The rates $x k$ and $\{p_s\}$ are always positive, so the bounds are zero and $2G_{ss}$. 
The real part of the non-zero eigenvalues of the matrix ${\bf G}[x, T]$ describe the rate at which the eigenvectors decay in the transient solution to equation \ref{eqn:time_derivative_y}. Note that the eigenvalues are functions of the mobile concentration $x$ and temperature $T$. 

\renewcommand{\arraystretch}{1.6}
\newcolumntype{K}[1]{>{\centering\arraybackslash}p{#1}}
\begin{table}[]
\centering
\begin{tabular}{|c|c|c|c|c|}
\hline 
\multirow{2}{*}{Energy (eV)} & \multicolumn{2}{c|}{W \cite{heinola-vacancy, heinola_tungsten_1}} & \multicolumn{2}{c|}{V (this work)}\tabularnewline
\cline{2-5} \cline{3-5} \cline{4-5} \cline{5-5} 
 &  & ZPE &  & ZPE \tabularnewline
\hline 
\hline
% \hline 
$E^b_{1}$ & 1.28 & 0.15 & 0.4 & 0.18\tabularnewline
% \hline 
$E^b_{2}$ & 1.25 & 0.16 & 0.49 & 0.17\tabularnewline
% \hline 
$E^b_{3}$ & 1.11 & 0.11 & 0.32 & 0.18\tabularnewline
% \hline 
$E^b_{4}$ & 1.00 & 0.11 & 0.3 & 0.18\tabularnewline
% \hline 
$E^b_{5}$ & 0.91 & 0.09 & 0.27 & 0.19\tabularnewline
% \hline 
$E^b_{6}$ & 0.32 & 0.15 & 0.17 & 0.15\tabularnewline
\hline 
$E^f(H_{\rm int})$ & 0.69 & 0.27 & -0.32 & 0.24 \tabularnewline
% \hline
$E_{m}$ & 0.21 & -0.04 & 0.07 & 0.03 \tabularnewline
\hline
\end{tabular}
\caption{Calculated values for the incremental binding energy of H to a H-vacancy complex $E^b_{i} = E^b_{H\rightarrow(i-1)H+vac}$, as well as H interstitial formation energy and the migration energy between adjacent \textit{tetrahedral} interstitial sites, using DFT data. The zero-point energy (ZPE) entries are corrections to the corresponding quantity.}
\label{table:dftinput}
\end{table}

%%%%%%%%%%%%%%%%%%%%%%%%%%%%%%%%%%%%%%%%%%%%%%%%%%%%%%%%%%%%%%%%%

\subsubsection{Density Functional Theory (DFT) calculations}\label{sec:theory:dynamic_ss:DFT}

\noindent
To investigate the rate of convergence to steady state, we construct ${\bf G}$ matrices parameterized by density functional theory for the cases of tungsten and vanadium. We model the monovacancy with a maximum occupancy of six gas atoms. The formation energy of H in an interstitial site and saddle site, as well as the formation energy of a monovacancy and the formation energies of each H-vacancy complex, were required. For W, the values from previous works are reported for comparison \cite{heinola-vacancy,heinola_tungsten_1}. For V, we performed DFT calculations using the \textit{Vienna Ab initio Simulation Package} ({\tt VASP}) code \cite{kresse0} with the projected augmented-wave (PAW) method \cite{Bloechl_PRB_1994_Projector,kresse1}, and using the generalized gradient approximation (GGA) exchange correlation (XC) functional by Perdew, Burke, and Ernzerhof (PBE) \cite{Perdew_PRL_1996_Generalized}. The Methfessel-Paxton smearing method \cite{methfessel89high} with a smearing width of 0.1\,eV was used to approximate the orbital occupation function. The calculations were performed on 128-atom supercells (modified accordingly for monovacancy and incremental H) with a plane-wave energy cut-off of 450\,eV and a 4$\times$4$\times$4 \textit{k}-point grid. \\

\noindent
The interstitial formation energy and saddle point formation energy are denoted as $E^f(H_{\rm int})$ and $E^f(H_{\rm sad})$ respectively. The $i$H-vacancy formation energy is given as $E^f(iH+{\rm vac})$. From these energies, the migration energy and incremental binding energies are calculated in accordance with the Heinola definition \cite{heinola-vacancy, heinola_tungsten_1} given in equations \ref{eqn:migration_energy_def} and \ref{eqn:binding_energy_def}. The interstitial formation energy and the incremental binding energies for V have been validated with values from literature~\cite{ohsawa,vanadium_dft}. The energy data used in this work is compiled in table~\ref{table:dftinput}. To complete the parameterization for the ${\bf G}$ matrix we use indicative placeholder values $\{\nu,g,g'\} = \{10^{13},1,6\}$.

\begin{equation}
    \label{eqn:migration_energy_def}
    \begin{gathered}
    E_m = E^f(H_{\rm sad}) - E^f(H_{\rm int})  
    % E^*_m = E^*_f(H_{sad}) - E^*_f(H_{int})
    \end{gathered}
\end{equation}

\begin{equation}
    \label{eqn:binding_energy_def}
    \begin{gathered}
    E^b_i = E^f((i-1)H+{\rm vac}) + E^f(H_{\rm int})-E^f(iH+\rm vac)  
    % E^*_i = E^*_f((i-1)H+vac) + E^*_f(H_{int})-E^*_f(iH+vac)
    \end{gathered}
\end{equation}

%%%%%%%%%%%%%%%%%%%%%%%%%%%%%%%%%%%%%%%%%%%%%%%%%%%%%%%%%%%%%%%%%%%

\subsubsection{Analytic justification for steady state}\label{theory:dynamic_ss:analytic}

\begin{figure}[ht]
    \subfigure[W monovacancies]{
        \includegraphics[width=1\linewidth]{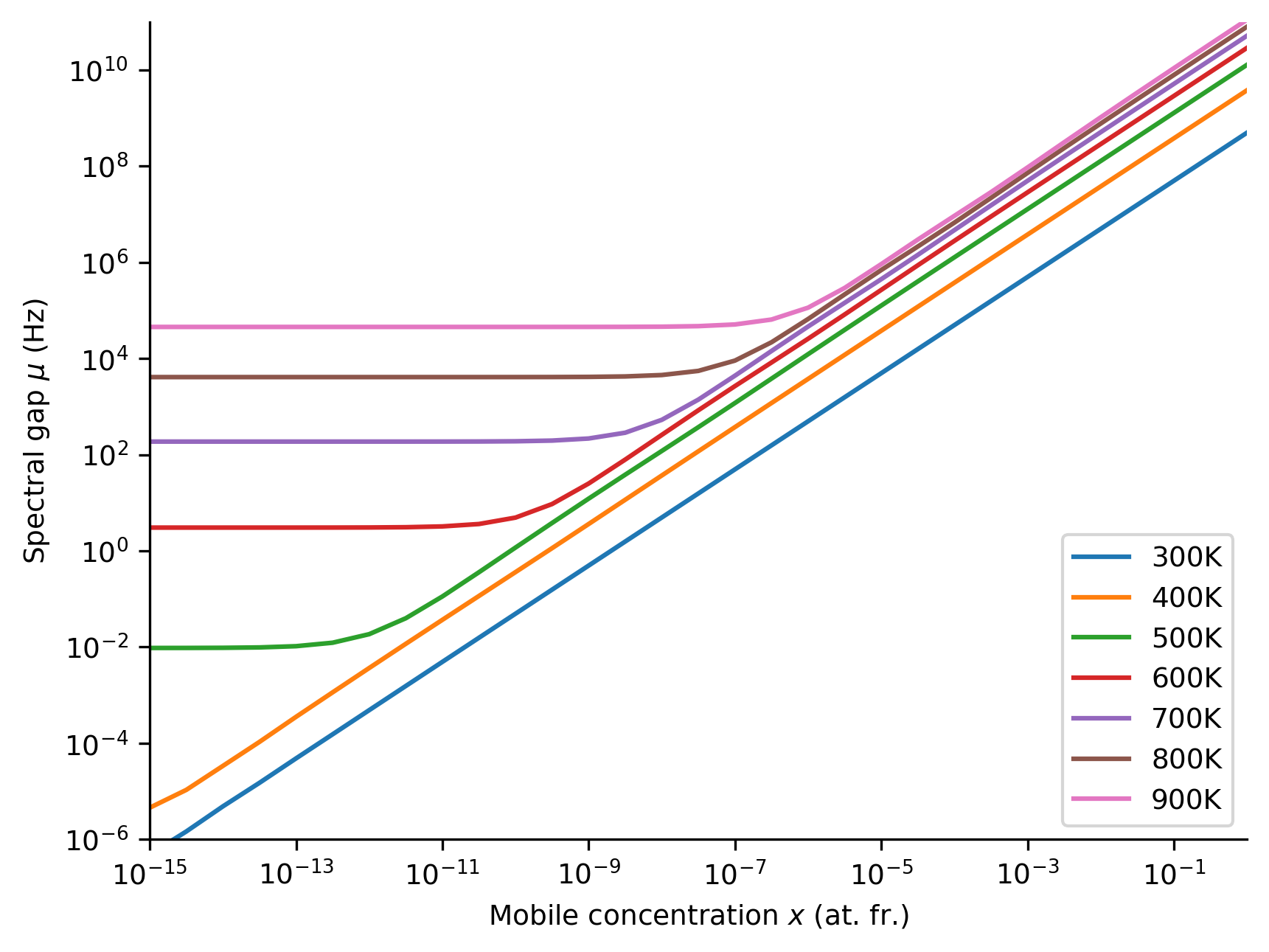}
        \label{fig:W_eigenvalues}
    }
    \subfigure[V monovacancies]{
        \includegraphics[width=1\linewidth]{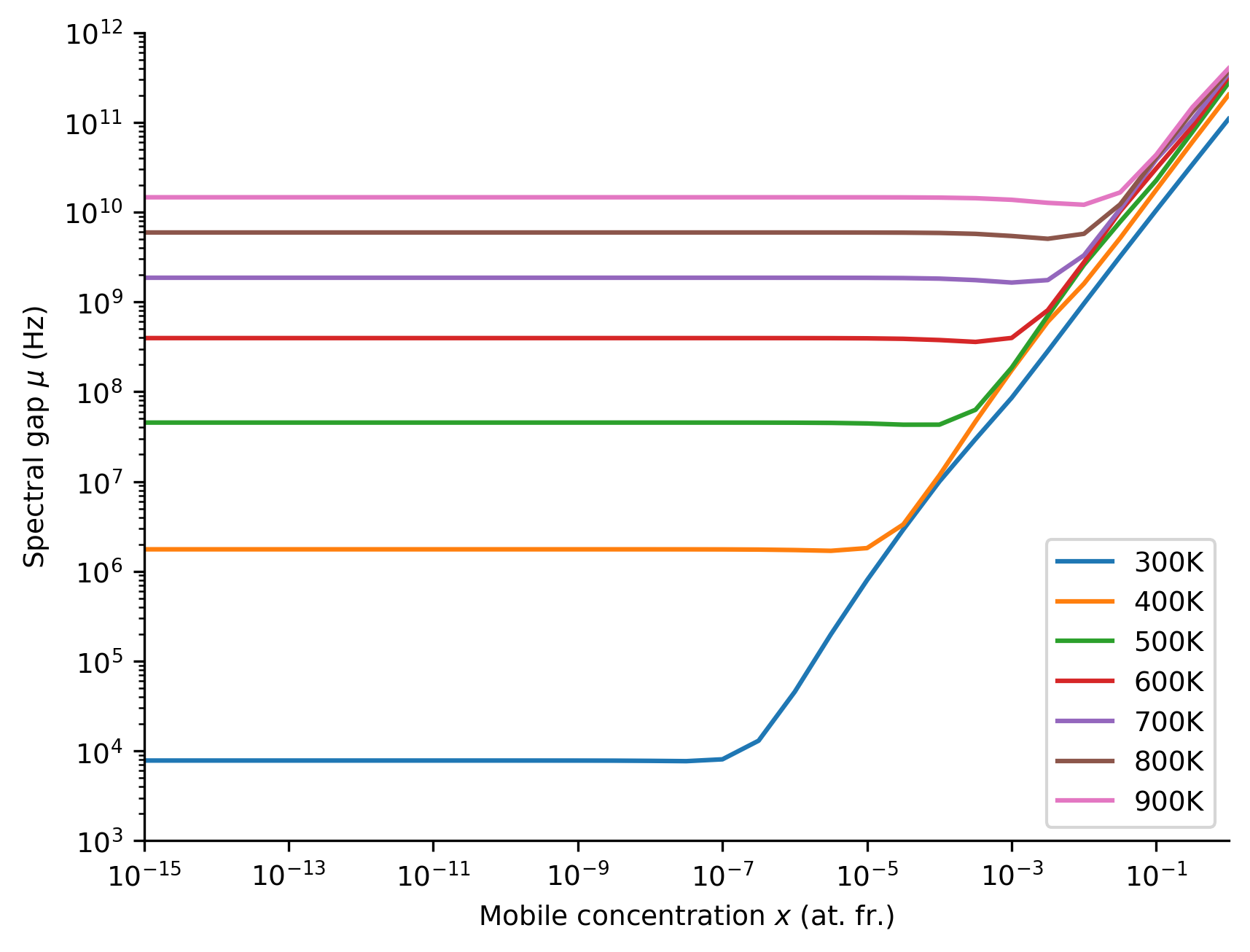}
        \label{fig:V_eigenvalues}
    }
    \caption{The ${\bf G}$ matrix spectral gap (Hz), the rate of convergence to steady state, at various mobile H concentrations and temperatures for W and V monovacancies.}
    \label{fig:spectralgap}
\end{figure}

\noindent
The spectral gap $\mu(x,T)$ is the real part of the smallest magnitude non-zero eigenvalue for the system and describes the \textit{slowest} rate of convergence to the steady state. The timescale of convergence is $\sim 1/\mu$. Figures \ref{fig:W_eigenvalues} and \ref{fig:V_eigenvalues} show the dependence of $\mu$ on $x$ and $T$, for W and V monovacancies respectively. We see as $x \rightarrow 0$, $\mu$ plateaus for a given $T$ so the spectral gap is strictly positive and the system always has a finite convergence rate. Gas loading and unloading correspond to increases and decreases in mobile concentration: using $\mu \propto x$, loading would support a quicker equilibration while unloading would continue to slow equilibration. \\

\noindent
As derived in the appendix \ref{sec:appendix:B}, the change in mobile concentration with time from sources or diffusion needs to be sufficiently small for the system to settle into steady state. If this condition does not hold, equation \ref{eqn:time_derivative_y} should be integrated in time and may be considered a reaction-limited regime. But if the condition does hold and the system indeed relaxes, we are in a diffusion limited regime. It can be shown that the steady state \textit{persists} with small, local changes in mobile concentration, see appendix \ref{sec:appendix:C}. \\

\noindent
For retention studies post-irradiation, the gas exposure period is in the order of hours. Figure \ref{fig:W_eigenvalues} shows that, for tungsten monovacancies, it is probably a fair assumption that the system reaches steady state in this time, unless under conditions of very low temperature and mobile gas concentration. For practical gas loading/unloading temperatures of 500K and above, it is reasonable to take the steady state. Figure \ref{fig:V_eigenvalues} suggests that it is always reasonable to model trapping in vanadium vacancies using the dynamic steady state.

%%%%%%%%%%%%%%%%%%%%%%%%%%%%%%%%%%%%%%%%%%%%%%%%%%%%%%%%%%%%%%%%%%%%%

\subsubsection{Effective diffusivity for a single equilibrated trap} \label{sec:theory:dynamic_ss:oriani}

\noindent
Oriani \cite{oriani} formalised an effective diffusivity in order to quantify the reduction in mobile gas diffusivity due to traps in a lattice. This was done by considering Fick's first law: the mobile gas flux is proportional to the mobile concentration gradient, ${\bf J}_x = -D \nabla x$. It was argued that the same flux drives the \textit{total} concentration gradient with an effective diffusivity $D_{\rm eff}$ that takes the equilibrated trapped concentration into account, i.e. ${\bf J}_c = -D_{\rm eff} \nabla c$.  The Oriani effective diffusivity $D^{\rm Oriani}_{\rm eff}$ was derived as the ratio of the spatial gradients of the mobile and total concentrations in the $z$-direction,

\begin{eqnarray}
    \label{eqn:oriani_Deff}
    D_{\rm eff}^{\rm Oriani} &=& D \frac{\partial x/dz}{\partial c/dz} = D \frac{\partial x}{\partial c} = D \left( \frac{\partial c}{\partial x} \right)^{-1} \nonumber\\
    &=& \left( 1 + \frac{\partial c_t}{\partial x} \right)^{-1} D.
\end{eqnarray}

\noindent
Schmid \textit{et al.} \cite{schmid-Deff} calculated the effective diffusivity with one McNabb-Foster trap across many temperatures and mobile concentrations in W using this form. The temperature and mobile concentration directly determines whether a trap is empty, partially filled or completely filled in its dynamic steady state. Traps will impede diffusion until they are completely filled, after which the remaining mobile concentration can no longer be trapped. \\

\noindent
We can derive the equivalent result for the multi-occupancy trap as follows. Ignoring sources and boundary conditions for now, the rate of change of total concentration must be due to the gradient of the flux of mobile gas, i. e.
    \begin{equation}
        \frac{\partial c}{\partial t} = - \nabla {\bf J}_x.
    \end{equation}
If there is one trap and it is equilibrated, then equation \ref{eqn:totalconc} gives
    \begin{eqnarray}
        \label{eqn:totalconc_timederiv}
        \frac{\partial c}{\partial t} = - \nabla {\bf J}_x&=& \frac{\partial x}{\partial t} + \rho \, {\bf C} \cdot \frac{ \partial \textbf{y}^{\rm eq}}{\partial t}  \\
        &=& \left( 1 + \rho \, {\bf C} \cdot \frac{ \partial \textbf{y}^{\rm eq}}{\partial x}\right) \frac{\partial x}{\partial t}  ,
    \end{eqnarray}
and if the terms in equation \ref{eqn:totalconc_timederiv} are not spatially varying we can define an effective diffusivity by $\frac{\partial x}{\partial t} = D_{\rm eff} \nabla^2 x$, where
    \begin{equation}
        \label{eqn:oriani_multi_Deff}
        D_{\rm eff} = \left( 1 + \rho \, C \cdot \frac{ \partial \textbf{y}^{\rm eq}}{\partial x}\right)^{-1} D.
    \end{equation}
Making the substitutions in equation \ref{eqn:Gsingleoccu} will reduce this expression to the Oriani effective diffusivity, equation \ref{eqn:oriani_Deff}, in the single occupancy case.
Some manipulation gives closed forms for the total concentration $c$ in terms of the mobile fraction $x$ and the effective diffusivity prefactor, defined by $A \equiv D_{\rm eff}/D$,
\begin{eqnarray}
    c &=& x \left( \frac{ \alpha + x + \rho}{\alpha + x} \right)    \nonumber\\
        \label{eqn:eff_diff_prefactor}
        A &\equiv& \left( 1 + \rho \, C \cdot \frac{ \partial \textbf{y}^{\rm eq}}{\partial x}\right)^{-1} \nonumber \\
        &=& \frac{ (c-\rho+\alpha) + \sqrt{ 4 c \alpha + (c - \rho - \alpha)^2} } { 2 \sqrt{ 4 c \alpha + (c - \rho - \alpha)^2} },
    \end{eqnarray}
where we have written the shorthand $\alpha = g' \exp\left[ - \frac{E^b}{k_BT}\right]$.
The effective diffusivity prefactor has a sigmoidal shape with the limits
    \begin{eqnarray}
    \label{eqn:Deff_limits}
        \lim_{c\rightarrow 0} A(c) &=& \frac{ \alpha} {\alpha + \rho}       \nonumber   \\
       \lim_{c\rightarrow \infty} A(c) &=& 1       \nonumber   \\
        A( c \approx \rho ) &=&  \frac{1}{2}\left( 1 + \frac{\alpha}{\sqrt{\alpha(\alpha+4\rho)}} \right)
        + \frac{2 \alpha \rho }{\sqrt{\alpha(\alpha+4\rho)}} (c - \rho) \nonumber \\
        &&+ \mathcal{O}\left( (c - \rho)^2 \right)  .   \nonumber \\
    \end{eqnarray}

\subsubsection{Diffusion and retention equations for multiple traps} 
\label{sec:theory:dynamic_ss:multi-dynamics}
\noindent
If there are multiple traps, some of which are equilibrated and others not, and we include source and boundary conditions, then equations \ref{eqn:multiocc-evolution} and \ref{eqn:totalconc_timederiv} lead to very general equations for mobile and trapped gas evolution:
    \begin{eqnarray}
        \label{eqn:totalconc_timederiv_general}       
           \frac{\partial x}{\partial t} &=& \left( 1 + \sum_{j \in \rm eq} \rho_j \, {\bf C}_j \cdot \frac{ \partial {\bf y}^{\rm eq}_j}{\partial x}\right)^{-1} \nonumber\\
           && \times \left( \nabla \cdot( D \nabla x )+ \left. \frac{\partial x}{\partial t} \right|_{\rm source,bc} + \sum_{j \notin \rm eq} \rho_j \, {\bf C}_j \cdot {\bf G}_j {\bf y}_j  \right)  \nonumber\\
        \frac{\partial {\bf y}_j }{\partial t} &=& - {\bf G}_j {\bf y}_j        \quad \quad  \quad j \notin \rm eq.
    \end{eqnarray}
 
%%%%%%%%%%%%%%%%%%%%%%%%%%%%%%%%%%%%%%%%%%%%%%%%%%%%%%%%%%%%%%%%%%%%%

\subsubsection{One n-occupancy trap vs n single occupancy traps} \label{sec:theory:dynamic_ss:diff_comparison}

\noindent
In this section, we compare the effective diffusivity with the multi-occupancy, incremental binding model to that calculated with multiple single occupancy traps. For a single isotope in a multi-occupancy trap, we can solve for the dynamic steady state analytically.
By defining a trapping-to-detrapping ratio $q_s = x k/p_s$ for each occupancy state $s$, the general $n$-occupancy $\textbf{y}^{\rm eq}$ vector can be written concisely as
\begin{equation}
\label{eqn:yeq_multiocc}
    \textbf{y}^{\rm eq} = \frac{1}{1+\sum_{k=1}^n \prod_{i=1}^k q_i} \left( \begin{array}{c}
          1 \\
          q_1 \\
          q_1 q_2 \\
          ...\\
          q_1...q_n
        \end{array}
          \right).
\end{equation}
From this, we deduce the multi-occupancy effective diffusivity prefactor for a single isotope is
\begin{equation}
\label{eqn:multi_Deff}
    A =\left( 1+\rho \sum_{i=1}^n i\frac{\partial}{\partial x} \left[ \frac{\prod_{m=1}^i q_m}{1+\sum_{k=1}^n \prod_{m=1}^k q_m} \right] \right)^{-1}.
\end{equation}

\noindent
We show in the appendix \ref{sec:appendix:A} that in general
\begin{equation}
    \label{eqn:variance}
    {\bf A} =  \left( 1 + \sum_{j \in \rm eq}\frac{\rho_j}{x} \, {\rm Var} ( {\bf y}^{\rm eq}_j ) \right)^{-1}.
\end{equation}

\noindent
Equation \ref{eqn:variance} allows extension to high-occupancy defects, including vacancy clusters and voids. \\

\noindent
Now consider $n$ single occupancy equilibrated traps, labelled by index $i=1,...,n$. The binding energy of the $i^{th}$ trap is set to $E^b_i$, the $i^{th}$ incremental binding energy to the $n$-occupancy trap. Each single occupancy trap is given the same density as the $n$-occupancy trap $\rho_i=\rho$, so the maximum trapped concentration is the same in both cases. The steady state probability for the $i$th single occupancy trap is
\begin{equation*}
\label{eqn:yeq_singleocc}
    {\bf y}^{\rm eq}_i = \frac{1}{1+q_i} \left( \begin{array}{c}
          1 \\
          q_i
        \end{array}
          \right) \quad \quad i=1,...,n
\end{equation*}
so the effective diffusivity across $n$ single occupancy traps is
\begin{equation}
\label{eqn:single_Deff}
    A =\left( 1+\rho\sum_{i=1}^n \frac{\partial}{\partial x} \left[ \frac{q_i}{1+q_i} \right]\right) ^{-1}.
\end{equation}

\noindent
In general, equations \ref{eqn:multi_Deff} and \ref{eqn:single_Deff} are not the same. In equation \ref{eqn:multi_Deff}, the sum to unity constraint on the probability vector as well as the products of $q_i$ in equation \ref{eqn:yeq_multiocc} ensure that as mobile concentration changes, the set of probabilities adjust together. Equation \ref{eqn:single_Deff} does not consider this. Even if the total concentration of trapped gas atoms is the same in both models, multiple single traps and multi-occupancy traps do not produce the same diffusive behaviour. In section \ref{sec:results:diff-multivssingle}, we show that the deviation between these equations can be significant. \\

\noindent
As an aside, we note that the form of equation \ref{eqn:yeq_multiocc} means that the steady state depends only on terms of the form
    \begin{equation*}
        q_s =  \frac{ x }{ g' C_s } \exp \left[ \frac{E^b_s}{k_BT} \right].
    \end{equation*}
It is noteworthy that the attempt frequency drops out. If there was no difference in zero-point energy between the different hydrogenic isotopes, as assumed in {\tt TESSIM-X}, then all isotopes would return the same dynamic steady state and effective diffusivity prefactor $A$.

%%%%%%%%%%%%%%%%%%%%%%%%%%%%%%%%%%%%%%%%%%%%%%%%%%%%%%%%%%%%%%%%%%%
%%%%%%%%%%%%%%%%%%%%%%%%%%%%%%%%%%%%%%%%%%%%%%%%%%%%%%%%%%%%%%%%%%%

\subsection{Multiple isotopes}\label{sec:theory:multiisotopes}

%%%%%%%%%%%%%%%%%%%%%%%%%%%%%%%%%%%%%%%%%%%%%%%%%%%%%%%%%%%%%%%%%%%

\subsubsection{Mathematical description}\label{sec:theory:multiisotopes:maths}

\noindent
For multiple gas species, the matrix ${\bf G}$ is a function of several mobile concentration fields, $x^{\alpha}(\textbf{r}, t)$, with unique trapping rates $x^{\alpha}k^{\alpha}$ and detrapping rates $p_s^{\alpha}$. We add a label to our counting vector in order to define the number of trapped atoms of type $\alpha$ as ${\bf C}^{\alpha} \cdot {\bf y}$.
\begin{eqnarray*}
    k^{\alpha}[T] &=& g \frac{D^{\alpha}[T]}{a^2}    \\
    p^{\alpha}_s[T] &=& g' {\bf C}^{\alpha}_s k^{\alpha}[T] \exp\left[ \frac{-E^{b , \alpha}_s}{k_B T} \right],
\end{eqnarray*}
where $D^{\alpha}$ is the diffusion constant for gas species $\alpha$. For a single isotope we could make a simple association between state label $s$ and occupation number, as the states could be labelled $s \in \{ 0,1,\ldots, n \}$. But for the two-occupancy case, we track the number of each type of atom in the trap, i. e. $s \in \{ 00,10,01,20,11,02,\ldots \}$. The number of distinct states for $m$ isotopes follows the sequence of $(m-1)$-simplex numbers: for a maximum occupancy $n$ we have ${\rm dim}(\textbf{y}) = (n+1)$ for $m=1$, $n(n+1)/2$ for $m=2$, $n(n+1)(n+2)/6$ for $m=3$ and so on. \\

\noindent
To find the total gas concentration of gas species $\alpha$, we add labels to equation \ref{eqn:totalconc}
    \begin{equation*}
        c^{\alpha}({\bf r},t) = x^{\alpha}({\bf r},t) + \rho({\bf r},t) \, {\bf C}^{\alpha} \cdot {\bf y}({\bf r},t),
    \end{equation*}
Therefore we write the complete time evolution equations for multi-gas, multi-trap, multi-occupancy with both equilibrated and non-equilibrated traps as equation \ref{eqn:totalconc_timederiv_general} for each gas type,
    \begin{eqnarray}
        \label{eqn:Palioxis}
        \sum_{\beta} \left( \delta_{\alpha \beta} + \sum_{j \in \rm eq} \rho_j \, {\bf C}^{\alpha}_{j} \cdot \frac{ \partial \textbf{y}^{\rm eq}_j}{\partial x^{\beta}} \right) \frac{\partial x^{\beta}}{\partial t} =  \quad \quad  \nonumber\\        
        \quad \quad
        \nabla \cdot (D^{\alpha} \nabla x^{\alpha}) + \left. \frac{\partial x^{\alpha} }{\partial t} \right|_{\rm source,bc} + \sum_{j \notin \rm eq} \rho_j \, {\bf C}^{\alpha}_j \cdot {\bf G}_j {\bf y}_j.      \nonumber\\
    \end{eqnarray}

\noindent
Equation \ref{eqn:Palioxis} is the principal result of the formalism in this paper, and its consequences are explored below. \\

\noindent
The {\tt PALIOXIS} library has been developed at UKAEA to compute the terms ${\bf C}_j \,,\, {\bf D} \,,\, {\bf G}_j$, and $\textbf{y}^{\rm eq}_j$ used in equation \ref{eqn:Palioxis}, starting from DFT data sheets similar to table \ref{table:dftinput}.
The solution for the time evolution of multiple isotopes in multi-occupancy traps, of which some are in dynamic equilibrium with the mobile gas, is implemented as a {\tt MOOSE} (Multiphysics Object-Oriented Simulation Environment) application \cite{giudicelli2024moose} which calls {\tt PALIOXIS}. The figures in this paper are generated with outputs from {\tt PALIOXIS} and the {\tt MOOSE} application except where noted otherwise. 

%%%%%%%%%%%%%%%%%%%%%%%%%%%%%%%%%%%%%%%%%%%%%%%%%%%%%%%%%%%%%%%%%%%

\subsubsection{Zero-point energy (ZPE) corrections} \label{sec:theory:multiisotopes:ZPE}

\noindent
When confined to interstitial or trap sites in a metal lattice, as shown in figure \ref{fig:three_level_trap}, hydrogen isotopes vibrate in quantised modes. While quantum effects on diffusivity such as tunnelling are not considered in this work, we do include zero-point energy corrections in both interstitial and trap sites. The corrections are listed in table \ref{table:dftinput}. First, the corrections on the interstitial formation energy and saddle point formation energy will produce an adjusted migration barrier $\tilde{E}^m$ according to equation \ref{eqn:migration_energy_def}. For H in W using the values reported in \cite{heinola-vacancy}, we find $\tilde{E}^m=0.17$eV. For H in V, after applying ZPE corrections from current DFT calculations, we find $\tilde{E}^m=0.1$eV. \\

\noindent
The zero-point energy corrections for D (or T) are calculated by scaling the zero-point energy corrections to the to the hydrogen formation energies $ZPE$$(H_{\rm int})$ and $ZPE$$(H_{\rm sad})$ with mass, multiplying each by $1/\sqrt{2}$ (or $1/\sqrt{3}$) before using equation \ref{eqn:migration_energy_def} with $E^{f,*}=E^f+ZPE$. The validity of this mass approximation for hydrogen isotopes in bcc metals is discussed in detail in \cite{chapman-iron-hydrogen}. Applying the approximation leads to $\tilde{E}^m_D=0.182$eV and $\tilde{E}^m_T=0.187$eV in W. In the same manner, each formation energy $E^f(iH+{\rm vac})$ is corrected before equation \ref{eqn:binding_energy_def} is used to produce $\tilde{E}^{b}_i$. Because ZPE is unique for each isotope, there is a difference in the binding energy of 1D opposed to 1T to a monovacancy at some occupancy. The type of each atom already trapped will also inform the binding energy, which distinguishes different occupation states in a multi-occupancy trap for multiple isotopes. \\

\begin{figure}
    \centering
    \includegraphics[width=1\linewidth]{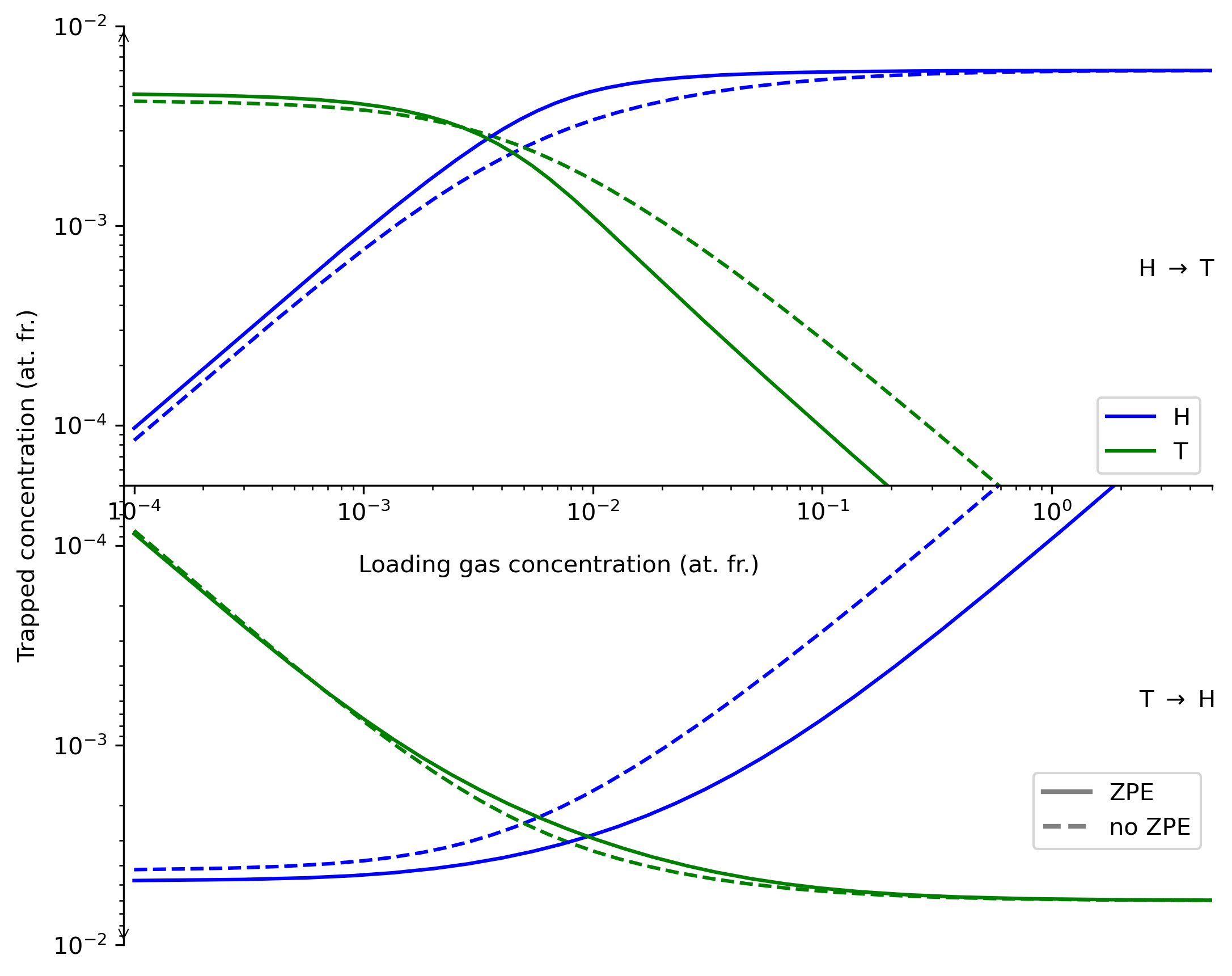}
    \caption{The trapped concentration of H and T in V monovacancies at 300K and density $\rho=10^{-3}$ at. fr. as a function of the loading gas concentration for H $\rightarrow$ T as well as T $\rightarrow$ H. Note the $y$-axis is reflected over $y=0.5 \times 10^{-4}$ at. fr.}
    \label{fig:V_HT}
\end{figure}

\noindent
We solved equations \ref{eqn:multiocc-evolution} in dynamic steady state with ZPE in order to investigate the variation in trapped gas with loading gas for H and T in V at 300K with six-occupancy monovacancies at trap density $10^{-3}$ at. fr. The results are presented in figure \ref{fig:V_HT}. The top plot considers H loading into V pre-loaded with T, initially stored as 5T-vacancy complexes when H concentration is low in comparison to trap density. As the H concentration increases, the trapped T is exchanged for H and once H is in excess, 6H-vacancy complexes dominate as a result of the low temperature. The bottom plot considers the opposite scenario, where most vacancies initially contain 5H then T is loaded. As expected, H is detrapped and exchanged for T. Without ZPE, H and T are described by the same ratio of trapping to detrapping so the scenarios perfectly mirror in behaviour. With ZPE however, we deduce that H loading flushes T out more significantly than the reverse process. We choose to illustrate this effect in V instead of W due to larger ZPE corrections associated with V monovacancies, given in table \ref{table:dftinput}.

%%%%%%%%%%%%%%%%%%%%%%%%%%%%%%%%%%%%%%%%%%%%%%%%%%%%%%%%%%%%%%%%%%%
%%%%%%%%%%%%%%%%%%%%%%%%%%%%%%%%%%%%%%%%%%%%%%%%%%%%%%%%%%%%%%%%%%%
%%%%%%%%%%%%%%%%%%%%%%%%%%%%%%%%%%%%%%%%%%%%%%%%%%%%%%%%%%%%%%%%%%%

\section{Results}\label{sec:results}

%%%%%%%%%%%%%%%%%%%%%%%%%%%%%%%%%%%%%%%%%%%%%%%%%%%%%%%%%%%%%%%%%%%
%%%%%%%%%%%%%%%%%%%%%%%%%%%%%%%%%%%%%%%%%%%%%%%%%%%%%%%%%%%%%%%%%%%

\subsection{H effective diffusivity against total concentration and temperature in W and V}\label{sec:results:diff-WvsV}

\begin{figure}
    \centering
    \subfigure[W monovacancies at density $\rho=10^{-3}$ at. fr.]{
        \includegraphics[width=1\linewidth]{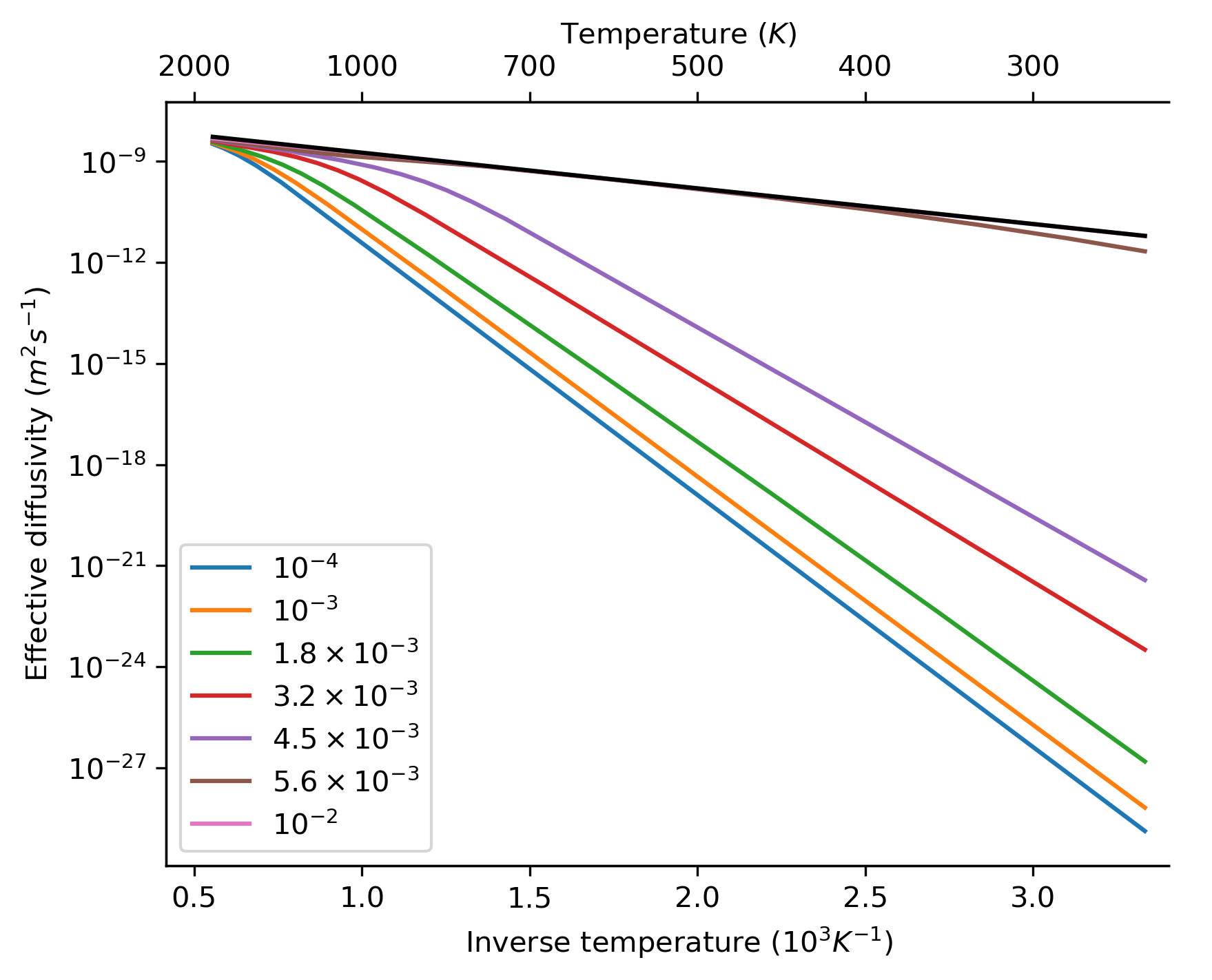}
        \label{fig:W_Deff_temp}
    }
    \subfigure[V monovacancies at density $\rho=10^{-3}$ at. fr.]{
        \includegraphics[width=1\linewidth]{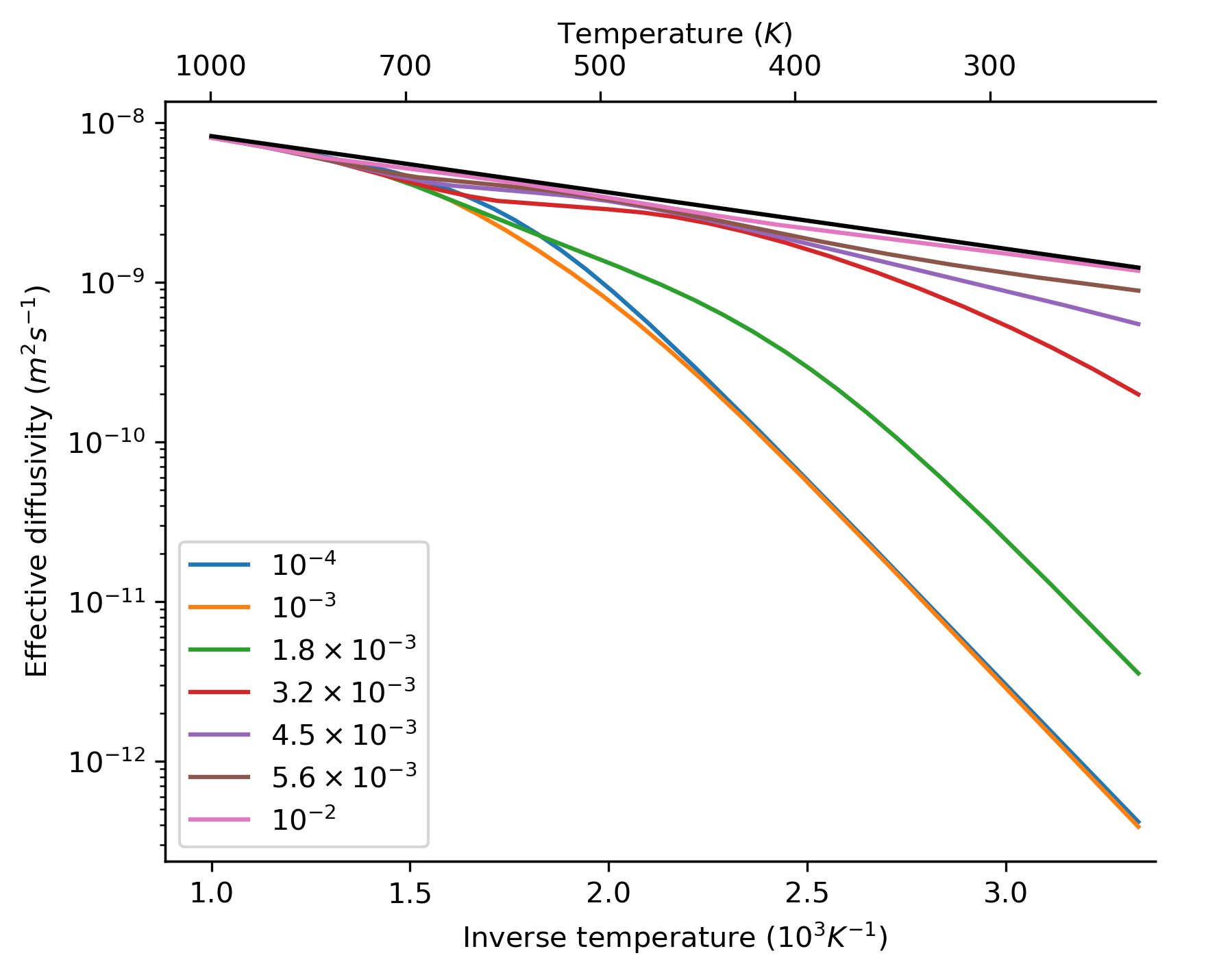}
        \label{fig:V_Deff_temp}
    }
    \label{fig:Deff_temp}
    \caption{The effective diffusivity of H in W and V, with equilibrated monovacancies at trap density $10^{-3}$ at. fr., as a function of inverse temperature across several total gas concentrations $c$. The black line denotes perfect lattice diffusivity.}
\end{figure}

\noindent
With equation \ref{eqn:multi_Deff} and the energies listed in table \ref{table:dftinput}, the effective diffusivity of H in both W and V with six-occupancy equilibrated monovacancies was computed for various temperatures and total gas concentrations, then plotted in figures \ref{fig:W_Deff_temp} and \ref{fig:V_Deff_temp}. The monovacancy density is fixed at $10^{-3}$ atomic fraction, a representative value for low-temperature irradiated materials \cite{Boleininger_SciRep2023}. A straight line in these Arrhenius plots with a gradient $d$ can be interpreted as a migration activation energy $E_a = -d$. For the perfect lattice we find $E_a = E_m$. These plots not only show that perfect lattice tungsten has a higher migration barrier than vanadium, hence H has a lower diffusivity in tungsten, but also show trapping and detrapping leads to non-Arrhenius behaviour. \\

\noindent
The effective diffusivity of H at total concentrations \textit{much less} than the trap density is much lower than the perfect lattice diffusivity, with a low temperature activation barrier $E_a \approx E_m + E^b_1$. This is because most traps are empty, and the small amount of mobile H will be trapped and detrapped as it diffuses. We expand on this match between single and multi-occupancy models at low gas concentrations later in the text. Once the total concentration far exceeds the trap density, the effective diffusivity tends to the perfect lattice diffusivity: most traps are full and the remaining mobile H cannot be trapped and detrapped as it diffuses. \\

\noindent
The solid lines in figure \ref{fig:single+multi} demonstrate this dependence of the effective diffusivity on the total H concentration at several temperatures for the same trap density. While the effective diffusivity in W monotonically increases with H concentration, V shows a dip in effective diffusivity for total concentrations close to the trap density. This behaviour is due to the incremental binding energies listed in table \ref{table:dftinput}: for V monovacancies $E^b_2 > E^b_1$ so 1H-vacancy complexes are more binding to a passing mobile H than empty vacancies. This is not the case in W, where the incremental binding energies are strictly decreasing. Vanadium is not an exception to have $E^b_{i+1} > E^b_{i}$, rather it is the norm: iron, chromium and tantulum monovacancies also demonstrate this non-monotonic behaviour \cite{ohsawa}.

%%%%%%%%%%%%%%%%%%%%%%%%%%%%%%%%%%%%%%%%%%%%%%%%%%%%%%%%%%%%%%%%%%%
%%%%%%%%%%%%%%%%%%%%%%%%%%%%%%%%%%%%%%%%%%%%%%%%%%%%%%%%%%%%%%%%%%%

\subsection{H diffusion and retention for 6-occupancy vs 6 single occupancy equilibrated traps} \label{sec:results:diff-multivssingle}

\begin{figure}
    \centering
    \subfigure[W monovacancies at density $\rho=10^{-3}$ at. fr.]{
        \includegraphics[width=1\linewidth]{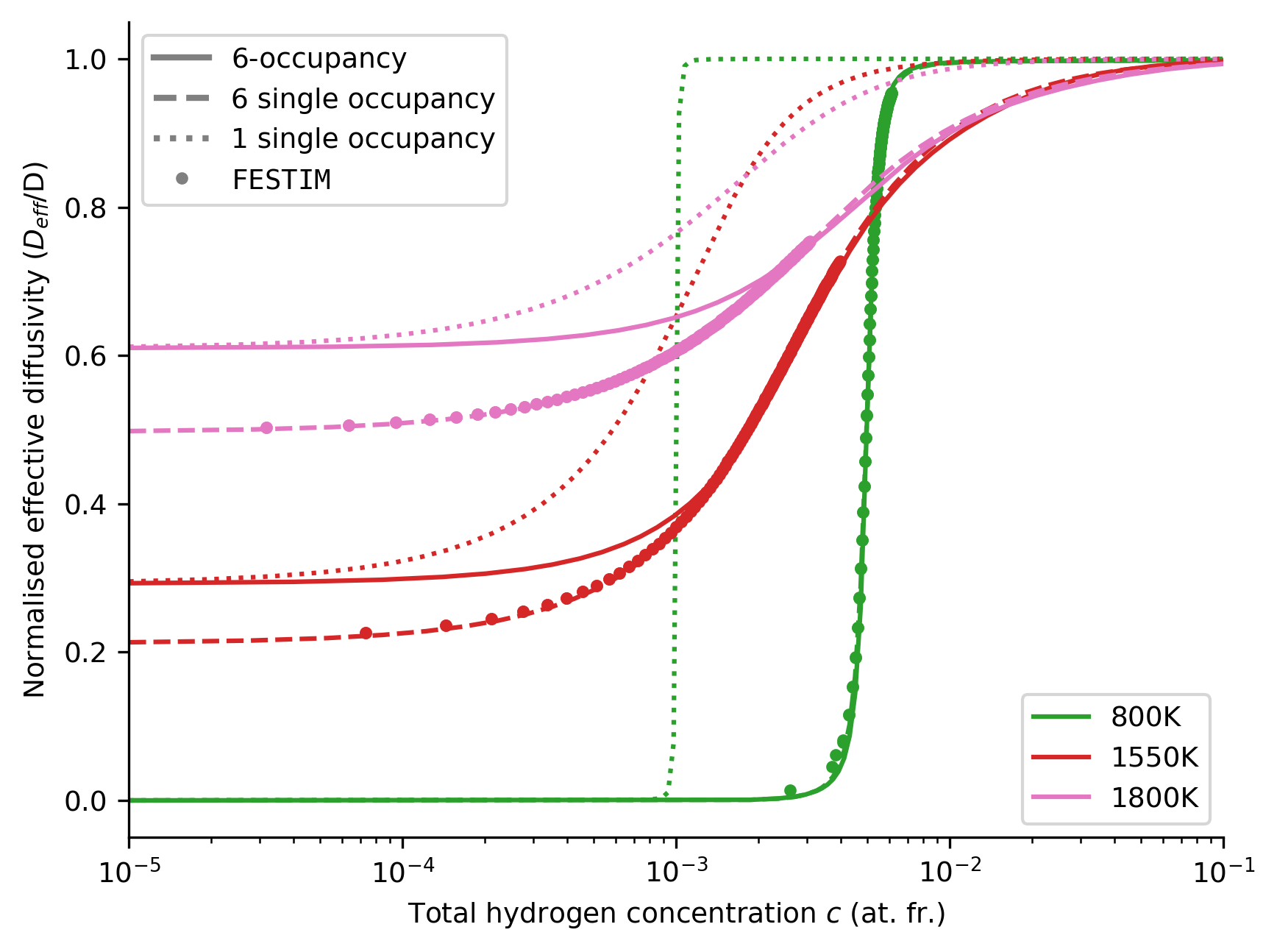}
        \label{fig:W_singlemulti}
    }
    \subfigure[V monovacancies at density $\rho=10^{-3}$ at. fr.]{
        \includegraphics[width=1\linewidth]{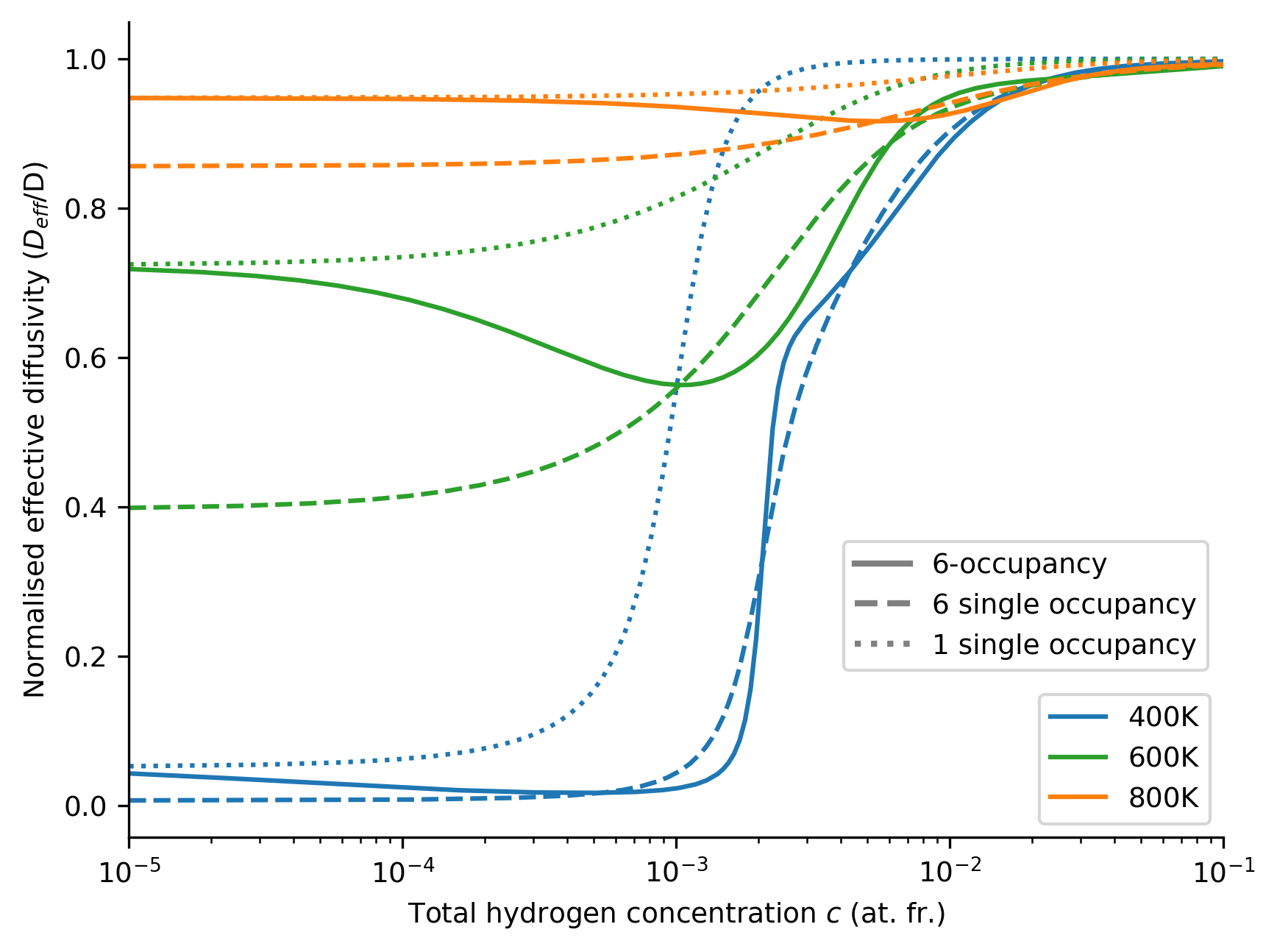}
        \label{fig:V_singlemulti}
    }
    \caption{Normalised effective diffusivity for H in W and V, with one six-occupancy equilibrated trap with incremental binding energies $\{E^b\}$; six single occupancy equilibrated traps labelled by index $i=1,...,n$ where the $i^{th}$ trap has binding energy $E^b_i$; and one single occupancy trap with $E^b_1$. The trap density is $10^{-3}$ at. fr. for all cases.}
    \label{fig:single+multi}
\end{figure}

\begin{figure}
    \centering
    \subfigure[W monovacancies at density $\rho=10^{-3}$ at. fr.]{
        \includegraphics[width=1\linewidth]{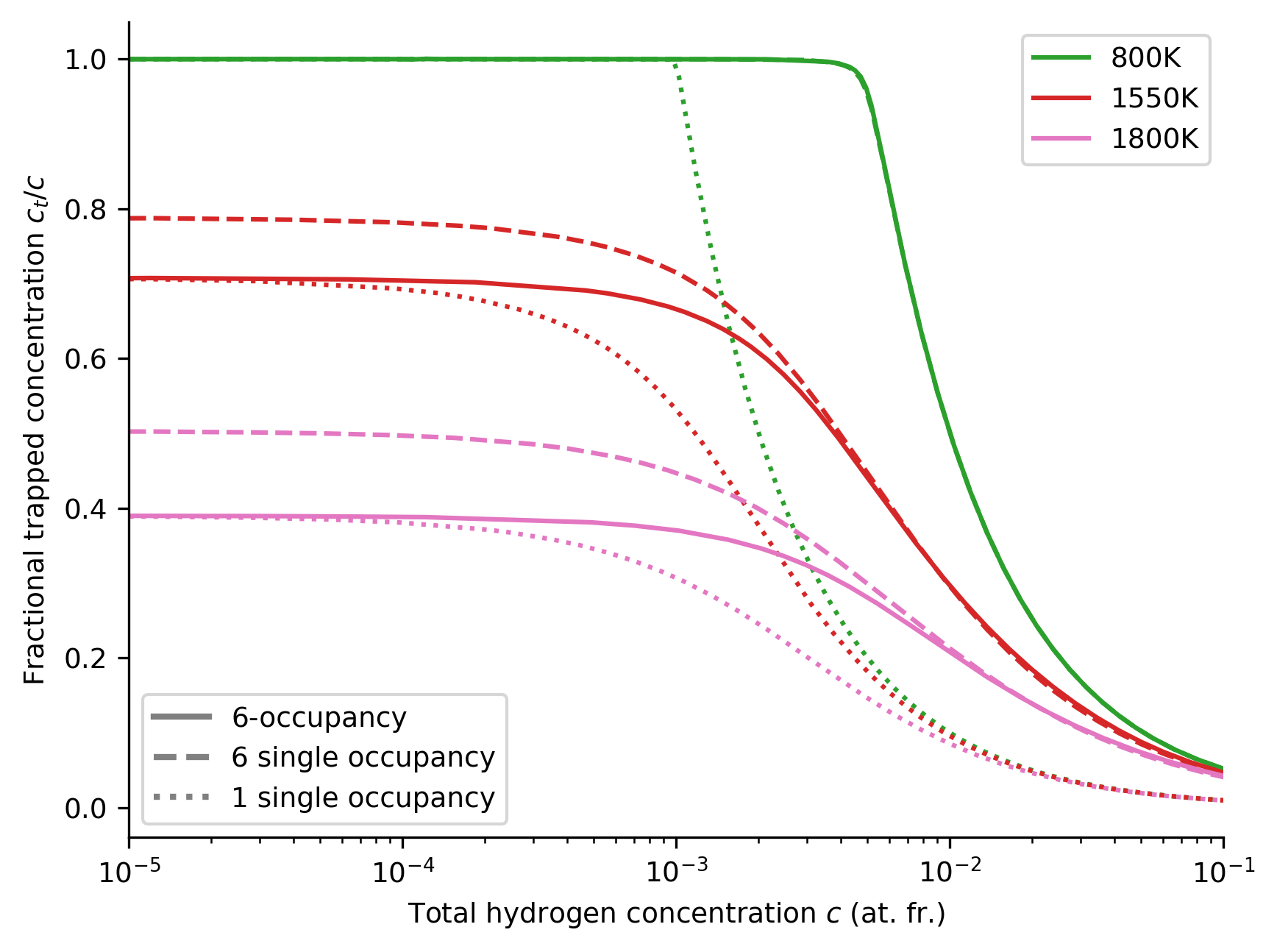}
        \label{fig:W_singlemulti}
    }
    \subfigure[V monovacancies at density $\rho=10^{-3}$ at. fr.]{
        \includegraphics[width=1\linewidth]{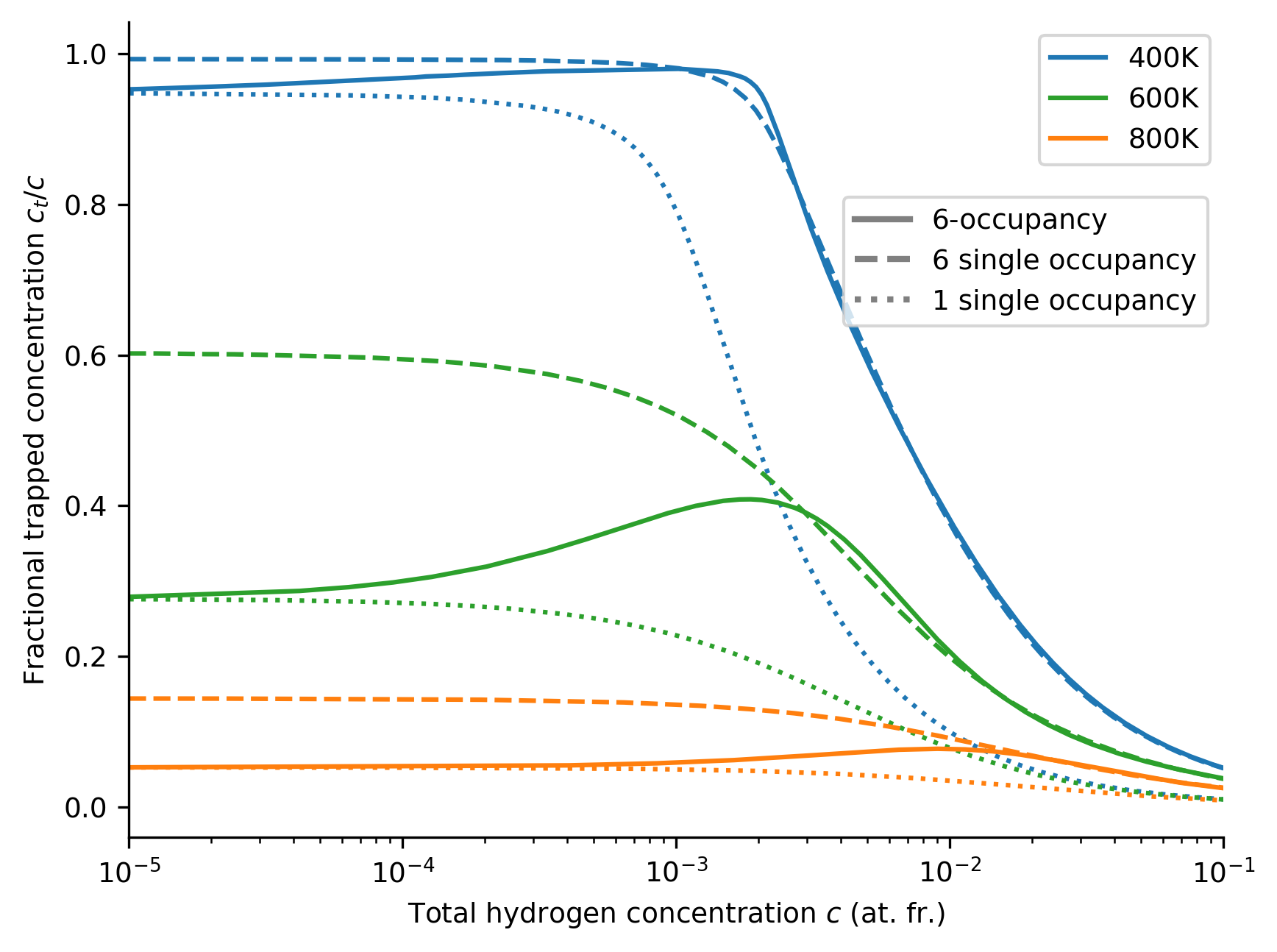}
        \label{fig:V_singlemulti}
    }
    \caption{The trapped H concentration $c_{t}$ as a fraction of total concentration $c$ in W and V, with one six-occupancy equilibrated trap with incremental binding energies $\{E^b\}$; six single occupancy equilibrated traps labelled by index $i=1,...,n$ where the $i^{th}$ trap has binding energy $E^b_i$; and one single occupancy trap with $E^b_1$. The trap density is $10^{-3}$ at. fr. for all cases.}
    \label{fig:single+multi_ret}
\end{figure}

\noindent
Figure \ref{fig:single+multi} presents the difference between the multi-occupancy effective diffusivity prefactor from equation \ref{eqn:multi_Deff} (solid lines) and the single occupancy equivalent from equation \ref{eqn:single_Deff} (dashed lines), parameterized by the binding energies in table \ref{table:dftinput} without ZPE. The single occupancy curves were produced for one trap with binding energy $E^b_1$, as well as six distinct traps corresponding to $\{E^b_i\}$. The six distinct trap curves were verified against transient {\tt FESTIM} calculations, by running several 1D simulations with this parameterization until dynamic steady state was achieved. Then, the Oriani effective diffusivity was computed using equation \ref{eqn:oriani_Deff}. Figure \ref{fig:single+multi_ret} presents the trapped concentration as computed by $\rho \textbf{C} \cdot \textbf{y}^\text{{eq}}$ per trap, taking the sum for six distinct single occupancy traps. All trap densities were set to $\rho=10^{-3}$ at. fr. \\

\noindent
Figures \ref{fig:single+multi} and \ref{fig:single+multi_ret} demonstrate how single occupancy traps differ from multi-occupancy traps in gas retention and effective diffusivity. We identify three critical regions in gas concentration relative to trap density. At low gas concentrations compared to trap density, traps are mostly empty or filled by one gas atom. The dotted lines, the effective diffusivity and retention with one single occupancy trap at $E^{b}_1$, match the solid lines in this region as a result. But multiple single occupancy traps do not have the same limit, as mobile gas atoms may be bound to any trap. \\

\noindent
At high gas concentrations where traps are mostly filled, all lines match diffusivity closely, as $A(c\rightarrow \infty) = 1$ for both equations \ref{eqn:multi_Deff} and \ref{eqn:single_Deff}. The multi-occupancy trap and six single occupancy traps also have the same total trapped retention, $c_t \sim 6 \rho$, while the one single occupancy trap has only $c_t \sim \rho$. In the intermediate region $c \sim \rho$, the effective diffusivity and gas retention are very sensitive to the trapping model used. This is the crucial point from an engineering perspective. During gas loading, the front indicating the depth of penetration will slowly advance, as the rising effective diffusivity allows gas to migrate deeper into the material only where the gas concentration reaches trap density. Consequently, the trapping model must be chosen carefully.

%%%%%%%%%%%%%%%%%%%%%%%%%%%%%%%%%%%%%%%%%%%%%%%%%%%%%%%%%%%%%%%%%%%
%%%%%%%%%%%%%%%%%%%%%%%%%%%%%%%%%%%%%%%%%%%%%%%%%%%%%%%%%%%%%%%%%%%

\subsection{Experimental comparison} \label{sec:results:experiment}

\noindent
Markelj \textit{et al.} \cite{nra_markelj} investigated H isotope exchange in polycrystalline tungsten damaged by 20MeV W$^{6+}$ ions. The sample was exposed to atomic H and D beams in sequence at 600K. Nuclear reaction analysis (NRA) \cite{nra-bulk-concentration} was used to determine D concentration as a function of depth on the order of micrometres at different time snapshots.
This experiment therefore shows the dynamic processes of isotope loading and exchange in irradiation damage defects in tungsten.
The experiment can be described in five stages, all at 600K. A) D loading for 48 hours (h), B) isothermal desorption for 43h, C) D loading for 24.5h, D) H loading for 96h, and  E) D loading for 71h. Stages D and E are the isotope exchange periods, as the stage before has already loaded the sample with the other isotope. \\

\noindent
Our intention is to model this experiment with equation \ref{eqn:Palioxis} without using fitting parameters to the experimental result, instead making a prediction from first principles.
First, the type and distribution of trapping sites $\rho_j({\bf r},t)$ through the sample should be treated. We considered three basic types of traps here - surface, bulk impurity sites, and irradiation-induced defects.
\begin{itemize}
    \item 
The surface sites are discussed in detail in Markelj \textit{et al.} \cite{nra_markelj}, including their possible role in isotope exchange via the Langmuir-Hinshelwood mechanism \cite{langmuir-hinshelwood}. Ogorodnikova \textit{et al.} have also modelled the effect of surface defect sites \cite{olga} on bulk gas retention, both \enquote{intrinsic} and ion-induced trap sites. In this work we assume that surface sites will show short-lived transients before becoming equilibrated, thus have little effect on the dynamics in the bulk. We therefore have ignored surface defects and expect not to reproduce the experimentally observed peak in gas retention seen a few tens of nanometres into the sample.
    \item 
We know that the material used in the experiment is not 100\% pure, so impurities will exist and have some effect on the dynamics~\cite{bulk-impurities}.
While we do not have a clear picture of what these impurities are, or what their binding energies should be, we can be reasonably sure that their density is small compared to the irradiation-induced defects.
    \item 
The irradiation-induced defects created by ion damage at 600K take the form of monovacancies and small vacancy clusters as well as dislocation loops \cite{X-Hu, X-Yi}.
We know that dislocation loops will not dominate, because a) they have a lower binding energy than vacancy defects \cite{mnl-potential, de-backer-disloc, terentyev-disloc} and b) there are many fewer dislocation core sites than vacancies, both on geometric grounds and due to sink bias \cite{disloc-sink}. Therefore we need only consider vacancy-type defects.
While Hou \textit{et al}. \cite{bcc-nanovoids} developed a good model for H binding to nanovoids, and their significance in 20MeV W irradiated samples is noted \cite{vac-clus}, here we restrict our attention to monovacancies for simplicity. We parameterize using Heinola's binding energies in table \ref{table:dftinput}, with $\{\nu, g,g'\}=\{10^{13}, 1,6\}$.
\end{itemize}

\begin{figure}
    \centering
    \includegraphics[width=1\linewidth]{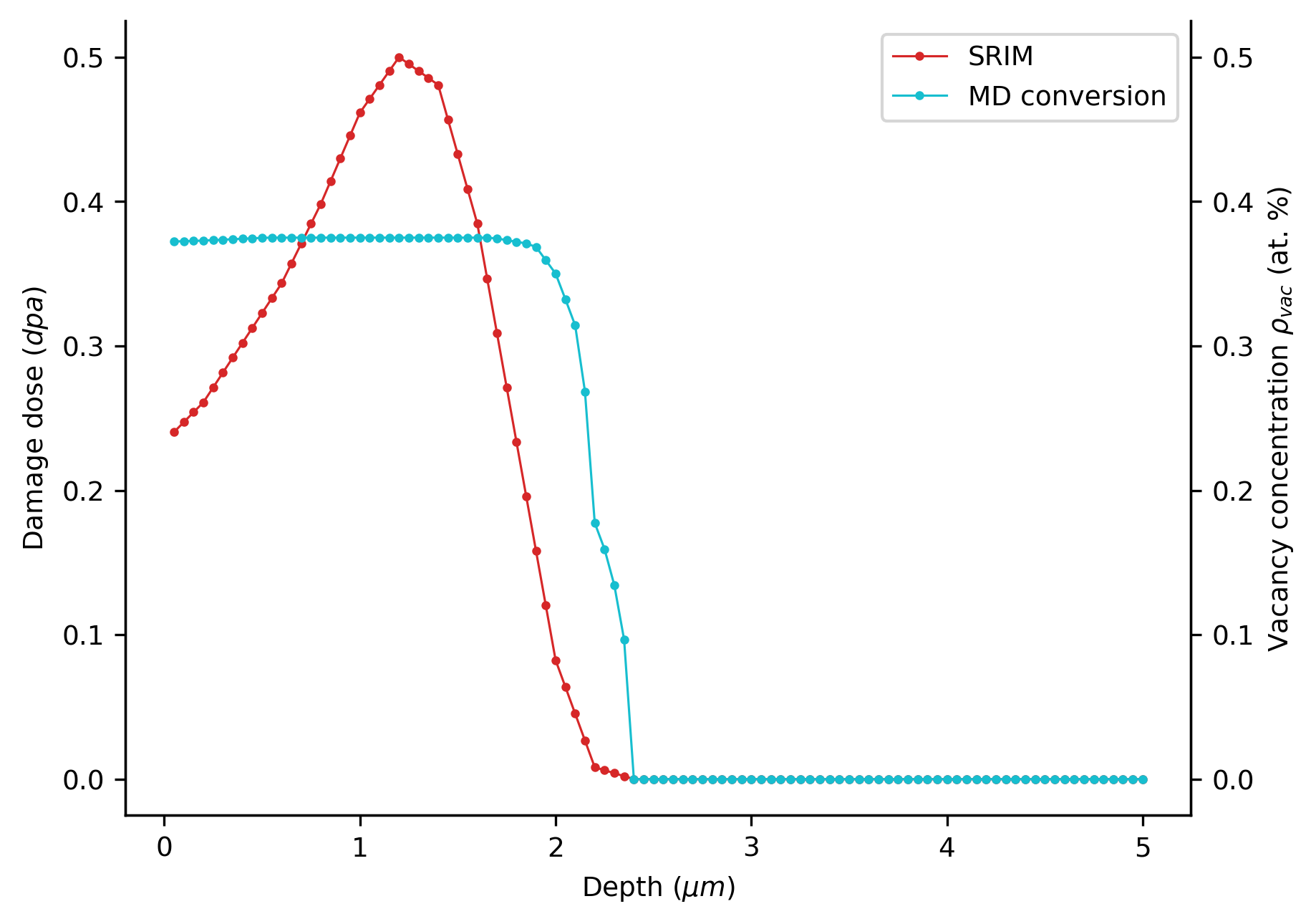}
    \caption{The damage-depth profile generated with the experiment parameters in {\tt SRIM}, and the corresponding monovacancy density profile predicted by direct MD cascade simulation at 300 K~\cite{md-conversion}.}
    \label{fig:srim_conversion}
\end{figure}

\noindent
We can estimate the distribution of vacancies using {\tt SRIM} \cite{SRIM} to produce a displacement per atom (dpa) against depth profile, then use molecular dynamics (MD) simulations to convert dpa into a monovacancy concentration \cite{md-conversion}. This is shown in figure \ref{fig:srim_conversion}: for damage above $0.1$dpa, the vacancy concentration saturates in the damaged region. We note that the MD simulations were performed at room temperature whereas the experiment is at 600 K, therefore we expect to somewhat \textit{overestimate} the vacancy concentration \cite{Markelj_2025}.\\

\noindent
The simulation in {\tt MOOSE} involved a one-dimensional mesh with length $L=$ 0.8mm. Gas loading was simulated by applying Dirichlet boundary conditions in mobile concentration, $x(z=0)=\kappa$ and $x(z=L)=0$ for the loading duration. The isothermal desorption in stage B was simulated with $x(z=0,L)=0$. We used the experimental fluxes of H and D, set at 6.9 $\times$ 10$^{18}$ m$^{-2}$s$^{-1}$ and 5.8 $\times$ 10$^{18}$ m$^{-2}$s$^{-1}$ respectively, and the implantation equation (labelled 20) in \cite{hodille-parameters}, with a beam implantation depth 6 nm and reflection coefficient 0.5, to calculate the constant source term $\kappa$. These parameters are order-of-magnitude estimates from similar implantation studies \cite{hodille-parameters}. \\

\noindent
The final part of the parameterization is the diffusion of gases in the perfect lattice, $D^{\alpha}$. For hydrogen, the hopping is between tetrahedral interstitial sites where $a = 1.11$ {\AA}. For deuterium, we use the mass scaling factor $1/\sqrt{2}$ to reduce the attempt frequency and zero-point energy on the migration barrier, as well as on the H-vacancy complex formation energies, as detailed in section \ref{sec:theory:multiisotopes:ZPE}. The {\tt PALIOXIS} library solved equation \ref{eqn:Palioxis} in each voxel, where the vacancies are treated as equilibrated and no other traps are considered.\\

\begin{figure}
    \centering
    \includegraphics[width=1\linewidth]{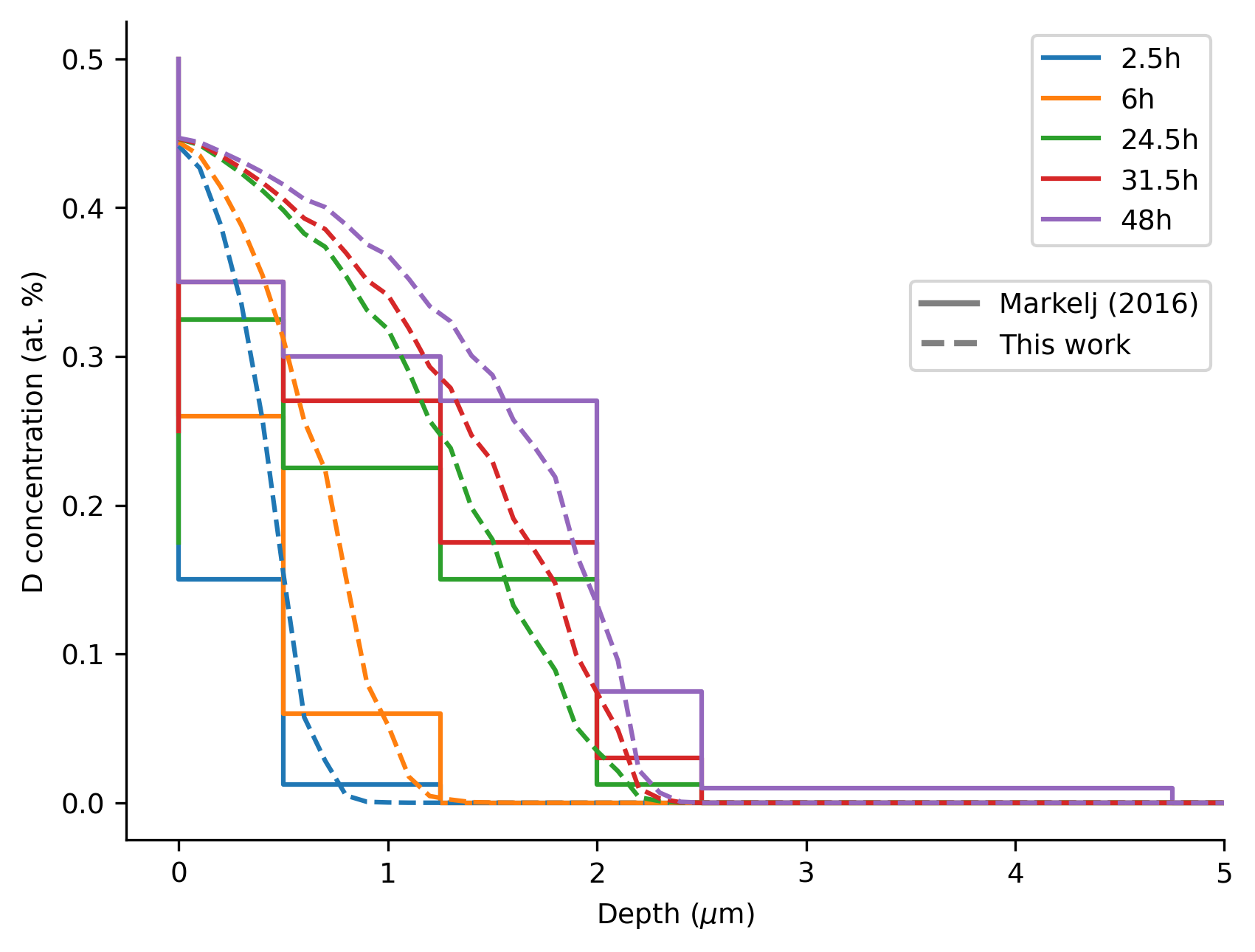}
    \caption{D concentration across irradiated sample depth during stage A) D loading for 48h, with simulation details outlined in section \ref{sec:results:experiment} and experimental data from \cite{nra_markelj}.}
    \label{fig:expcycle1}
\end{figure}

\begin{figure}
    \centering
    \includegraphics[width=1\linewidth]{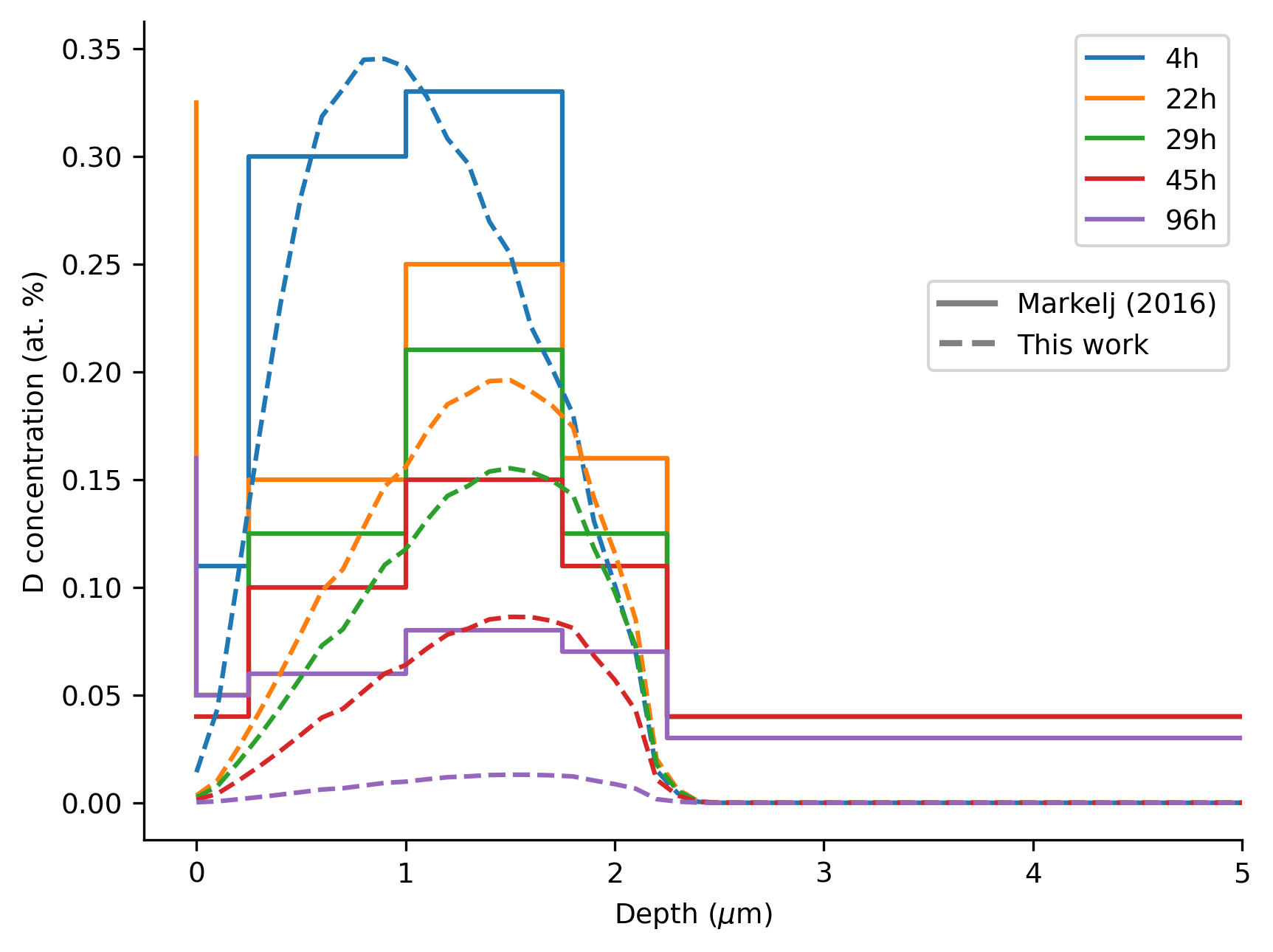}
    \caption{D concentration across irradiated sample depth during stage D) H loading for 96h, with simulation details outlined in section \ref{sec:results:experiment} and experimental data from \cite{nra_markelj}.}
    \label{fig:expcycle2}
\end{figure}

\noindent
The experimental and simulated D concentration-depth profiles for stages A and D are given in figures \ref{fig:expcycle1} and \ref{fig:expcycle2} respectively. The simulated retained gas is an overestimate, indicating the estimated monovacancy density was greater than the observed density. This suggests that work to improve parameter-free estimations of vacancy-type defect generation by irradiation at intermediate temperatures, 300 K $\le T \le$ 800 K, would be beneficial. We are also missing the narrow surface peak as expected. However, we see good agreement in the bulk between the computed and measured distributions in shape considering only a vacancy distribution is provided. In figure \ref{fig:expcycle2}, the measured D concentration beyond $2.5\mu$m is non-zero and homogeneous. This suggests that the H loading causes some D to diffuse deeper into the sample and become trapped by other defects, most likely impurities. \\

\begin{figure}
    \centering
    \includegraphics[width=1\linewidth]{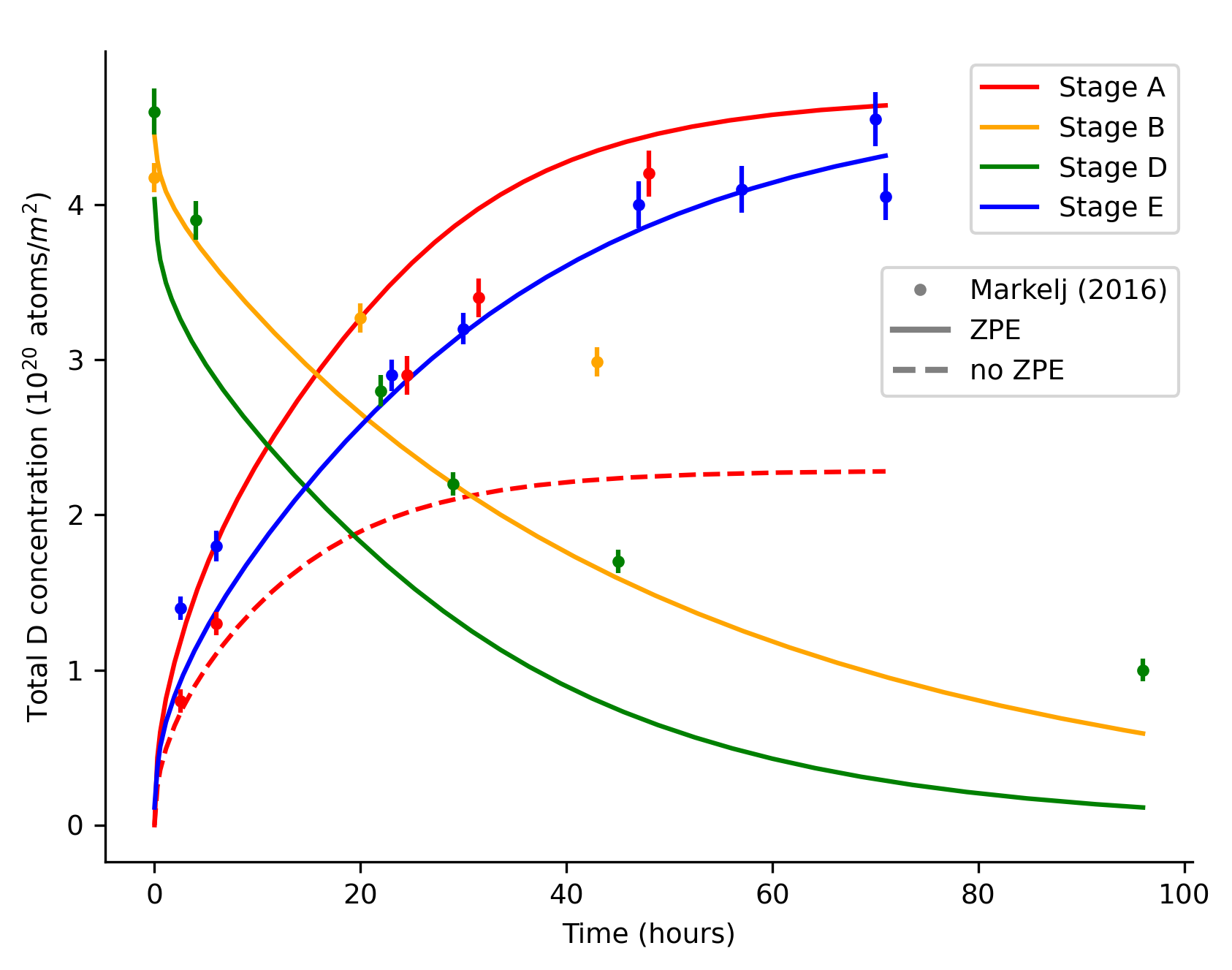}
    \caption{The total concentration of D during stages A, B, D and E of the experiment \cite{nra_markelj}, compared to those simulated in this work. The significance of ZPE on the simulated gas retention is also presented.}
    \label{fig:conc-time}
\end{figure}

\noindent
Figure \ref{fig:conc-time} presents the total D concentration over time during stages A and B, as well as the isotope exchange periods stages D and E. We see good agreement between the measured points and simulated curves during D uptake (blue and red), but some deviations in D release (green and yellow). Within the first 10 hours, the D uptake was measured to be greater in the pre-loaded H sample opposed to the empty sample, but we simulated no such initial increase in uptake. The uptake periods were also not measured to be as distinct as the simulated curves. The dashed curve is stage A) D loading for 48h, with \textit{no} zero-point energy corrections. Each incremental binding energy to the trap is lower without these corrections, so the dynamic steady state predicts less retention at the same temperature. Both simulated and experimental results show H loading in stage D flushes out D quicker than the isothermal desorption in stage B. \\

\noindent
More D is measured in experiment than predicted in both D release periods (green and yellow). This is consistent with impurity traps being present in experiment, but not modelled here. Ref \cite{nra_markelj} reports a second experimental set of total D concentration measurements during H loading in stage D, measured 2mm away from the coinciding H implantation beam and $^3$He measurement beam. We note this second set is a closer match to the simulated green curve, supporting our conclusion that another trap is responsible for the excess D measured, possibly the helium accumulated during the \textit{in situ} NRA measurements.

%%%%%%%%%%%%%%%%%%%%%%%%%%%%%%%%%%%%%%%%%%%%%%%%%%%%%%%%%%%%%%%%%%%
%%%%%%%%%%%%%%%%%%%%%%%%%%%%%%%%%%%%%%%%%%%%%%%%%%%%%%%%%%%%%%%%%%%
%%%%%%%%%%%%%%%%%%%%%%%%%%%%%%%%%%%%%%%%%%%%%%%%%%%%%%%%%%%%%%%%%%%

\section{Conclusions} \label{sec:conclusions}

\noindent
In this work, the mathematical formalism for multi-isotope retention and diffusion in multi-occupancy traps is described. When applied to H diffusion across W and V monovacancies, we have showed using DFT calculations that the trap dynamics are fast and motivate the use of a dynamic steady state. In the dynamic steady state, the probability vector for trap occupancies up to $n$ are uniquely defined, which then uniquely defines the trapped gas concentration and \textit{effective} diffusivity $D_{\rm eff}$. We show the dynamic steady state, hence effective diffusivity, for single and multi-occupancy traps are not the same in general, and may or may replicate the same retention and diffusive behaviour depending on total gas concentration. \\

\noindent
The effective diffusivity behaves differently in W and V with respect to total gas concentration. W reveals monotonically decreasing incremental H binding energies to its monovacancy, so $D_{\rm eff}$ monotonically increases as total H concentration rises. However the incremental binding energies to a V monovacancy do not follow this pattern, so there is a minimum in $D_{\rm eff}$ as total concentration approaches trap density. The total concentration at which $D_{\rm eff}$ rises or drops directly changes when gas may migrate deeper into the bulk, thus affecting long-term estimates of retention. \\

\noindent
For multiple isotopes, the model considers all unique occupancy states associated with $m$ isotopes in an $n$-occupancy trap using the simplex numbers. The detrapping rate for each isotope and from each occupancy state is \textit{unique}. The dynamic steady state vector is still one unique vector, but now depends on all mobile concentrations. We conclude the ZPE corrections on the migration barrier and the incremental binding energies for each isotope may significantly affect retention estimates, as shown for H and T binding to V monovacancies at room temperature. \\

\noindent
We have outlined a direct route from DFT calculations to gas retention estimates, with the appropriate statistical mechanics to account for sequential gas binding to traps. Only a vacancy distribution estimated with molecular dynamics is used as input to the model in order to replicate long-term retention measurements from a previous experiment. This supports the model's practicality for upscaling to reactor-scale investigations and tritium inventory modelling. Strongly binding, high occupancy traps like vacancy clusters and voids are abundant under fusion conditions. While not considered in this work, they can be modelled consistently with the use of equation \ref{eqn:variance} and multiple trapping sites. Consequently, the mathematical framework is easy to extend systematically in order to account for surface trapping or bulk impurities, through the use of distinct trap dynamics matrices to describe each trap type.

%%%%%%%%%%%%%%%%%%%%%%%%%%%%%%%%%%%%%%%%%%%%%%%%%%%%%%%%%%%%%%%%%%%
%%%%%%%%%%%%%%%%%%%%%%%%%%%%%%%%%%%%%%%%%%%%%%%%%%%%%%%%%%%%%%%%%%%
%%%%%%%%%%%%%%%%%%%%%%%%%%%%%%%%%%%%%%%%%%%%%%%%%%%%%%%%%%%%%%%%%%%

\section*{Acknowledgements} \label{sec:acknowledgements}

\noindent
This work has been carried out within the framework of the EUROfusion Consortium, funded by the European Union via the Euratom Research and Training Programme (Grant Agreement No 101052200 — EUROfusion) and from the EPSRC [grant number EP/W006839/1].  To obtain further information on the data and models underlying this paper please contact PublicationsManager@ukaea.uk. Views and opinions expressed are however those of the author(s) only and do not necessarily reflect those of the European Union or the European Commission. Neither the European Union nor the European Commission can be held responsible for them. \\

\noindent
This work has been (part-) funded by the EPSRC Energy Programme [grant number EP/W006839/1].  To obtain further information on the data and models underlying this paper please contact PublicationsManager@ukaea.uk.  \\

\noindent
The authors would like to thank Ed Shelton, Rishit Dhoot, Helen Brooks, and Steven Van Boxel for stimulating discussions in the development and validation of our model, and Kamran Pentland for a review of the mathematical analysis.

%%%%%%%%%%%%%%%%%%%%%%%%%%%%%%%%%%%%%%%%%%%%%%%%%%%%%%%%%%%%%%%%%%%
%%%%%%%%%%%%%%%%%%%%%%%%%%%%%%%%%%%%%%%%%%%%%%%%%%%%%%%%%%%%%%%%%%%
%%%%%%%%%%%%%%%%%%%%%%%%%%%%%%%%%%%%%%%%%%%%%%%%%%%%%%%%%%%%%%%%%%%

\section*{Data availability}\label{sec:data_availability}

\noindent
The data that support the findings of this article are openly available \cite{zenodo_src}. \\

%%%%%%%%%%%%%%%%%%%%%%%%%%%%%%%%%%%%%%%%%%%%%%%%%%%%%%%%%%%%%%%%%%%
%%%%%%%%%%%%%%%%%%%%%%%%%%%%%%%%%%%%%%%%%%%%%%%%%%%%%%%%%%%%%%%%%%%
%%%%%%%%%%%%%%%%%%%%%%%%%%%%%%%%%%%%%%%%%%%%%%%%%%%%%%%%%%%%%%%%%%%

\section*{CRediT}\label{sec:CRediT}

\noindent
{\bf S. Kaur} Methodology, Validation, Formal Analysis, Investigation, Writing,
{\bf D. Mason} Conceptualisation, Methodology, Software, Validation, Formal Analysis, Writing, Supervision,
{\bf S. Dixon} Software, Validation,
{\bf S. Mungale} Software, Validation,
{\bf P. Srinivasan} Investigation,
{\bf T. Orr} Investigation, Validation,
{\bf M. Lavrentiev} Supervision,
{\bf D.  Nguyen-Manh} Conceptualisation.

%%%%%%%%%%%%%%%%%%%%%%%%%%%%%%%%%%%%%%%%%%%%%%%%%%%%%%%%%%%%%%%%%%%
%%%%%%%%%%%%%%%%%%%%%%%%%%%%%%%%%%%%%%%%%%%%%%%%%%%%%%%%%%%%%%%%%%%
%%%%%%%%%%%%%%%%%%%%%%%%%%%%%%%%%%%%%%%%%%%%%%%%%%%%%%%%%%%%%%%%%%%

\section*{Declarations}\label{sec:declarations}

\noindent
The authors have no conflicts of interest to declare. 

%%%%%%%%%%%%%%%%%%%%%%%%%%%%%%%%%%%%%%%%%%%%%%%%%%%%%%%%%%%%%%%%%%%
%%%%%%%%%%%%%%%%%%%%%%%%%%%%%%%%%%%%%%%%%%%%%%%%%%%%%%%%%%%%%%%%%%%
%%%%%%%%%%%%%%%%%%%%%%%%%%%%%%%%%%%%%%%%%%%%%%%%%%%%%%%%%%%%%%%%%%%

\section{Appendix} \label{sec:appendix}

%%%%%%%%%%%%%%%%%%%%%%%%%%%%%%%%%%%%%%%%%%%%%%%%%%%%%%%%%%%%%%%%%%%
%%%%%%%%%%%%%%%%%%%%%%%%%%%%%%%%%%%%%%%%%%%%%%%%%%%%%%%%%%%%%%%%%%%

\subsection{The numerical method for finding the steady state eigenvector}\label{sec:appendix:A}

\noindent
The matrix ${\bf G}$ containing trapping and detrapping rates is very ill-conditioned, especially at low temperatures. From equation \ref{eqn:tridiagG}, Gershgorin's circle theorem~\cite{Gershgorin_mathworld} implies the ratio between largest and smallest non-zero eigenvalues will be of order $u_1/u_n = \exp\left[ (E_n-E_1)/(k_B T) \right]$. There is also a zero eigenmode because particle conservation requires the sum of all trapping and detrapping rates from a given occupancy state to be zero, i. e. $\sum_k {G}_{ki} = 0$. Therefore ${\bf G}$ has incomplete rank, as one row must be linearly dependent on the others, so there exists a vector ${\bf y}^{\rm eq}$ for which ${\bf Gy}^{\rm eq} = {\bf 0}$ and we name the dynamic steady state. \\

\noindent
To solve for ${\bf y}^{\rm eq}$, we precondition and symmetrise. Writing the preconditioning matrix $\textbf{N}$ as
    \begin{equation}
        {N}_{ij} = \delta_{ij} \left( {G}^T {G} \right)_{ii}^{-1/2},
    \end{equation}
we construct the better-conditioned symmetric matrix $\tilde{\bf G}$, given by
    \begin{equation}
        \tilde{\bf G} = ({\bf G}{\bf N})^T ({\bf G}{\bf N}),
    \end{equation}
then solve for the eigendecomposition of $\tilde{\bf G}$ using the {\tt LAPACK} routine {\tt DSYEV}~\cite{Lapack}.
From this, we find one zero eigenmode $\tilde{\bf G} {\bf z}^0 = {\bf 0}$, and from this recover
    \begin{equation}
        {\bf y}^{\rm eq} = {\bf N} {\bf z}^0.
    \end{equation}

\noindent
The derivative of gas retention in the dynamic steady state ${\bf C} \cdot {\bf y}^{\rm eq}$ with respect to mobile gas fraction ${\bf x}$ is found from the variance of the retention as follows. For a single isotope, we start with equation \ref{eqn:yeq_multiocc}. For occupancy state $s$, 
    \begin{equation*}
        \textbf{y}^{\rm eq}_s = \frac{\prod_{k=1}^{s} q_k}{1+\sum_{k=1}^n \prod_{i=1}^k q_i}. 
    \end{equation*}
To take the derivative with respect to the scalar mobile fraction $x$, we note $\frac{\partial q_s}{\partial x} = \frac{q_s}{x}$ for $s>0$, hence  
    \begin{equation*}
        \frac{\partial }{\partial x} y^{\rm eq}_s = \frac{s}{x} y^{\rm eq}_s - \sum_{k=1}^{s} \frac{k}{x} y^{\rm eq}_k y^{\rm eq}_s.
    \end{equation*}
From this we find
    \begin{equation*}
        \frac{\partial }{\partial x} {\bf C} \cdot {\bf y}^{\rm eq} 
        = \sum_s s \frac{\partial }{\partial x} y^{\rm eq}_s 
        = \frac{1}{x} \left( \sum_s s^2 y^{\rm eq}_s - \left(\sum_s s y^{\rm eq}_s\right)^2 \right),
    \end{equation*}
and conclude
    \begin{equation*}
        {\bf A} = \left( 1 + \rho \, {\bf C} \cdot \frac{ \partial \textbf{y}^{\rm eq}}{\partial x}\right)^{-1} 
       = \left( 1 + \frac{\rho}{x} \, {\rm Var} ( {\bf y}^{\rm eq} ) \right)^{-1}.
    \end{equation*}
As the expressions above are linear in trap density, for multiple traps we take the sum as in section \ref{sec:theory:dynamic_ss:multi-dynamics} to give
    \begin{equation}
        {\bf A} =  \left( 1 + \sum_{j \in \rm eq}\frac{\rho_j}{x} \, {\rm Var} (  {\bf y}^{\rm eq}_j ) \right)^{-1}.
    \end{equation}

\noindent
For multiple isotopes, a similar process gives us the effective diffusivity prefactor as the covariance,
    \begin{eqnarray}
        {\bf A}^{-1}_{\alpha \beta} 
        &=& \left( \delta_{\alpha \beta} +  \sum_{j \in \rm eq}\rho_j \,{\bf C}^{\alpha}_j \cdot \frac{ \partial \textbf{y}^{\rm eq}_j}{\partial x^{\beta}}\right)   \nonumber\\
        &=&  \left( \delta_{\alpha \beta} + \sum_{j \in \rm eq} \frac{\rho_j} {x^{\beta}} \,  {\rm Covar}_{\alpha \beta} (  {\bf y}^{\rm eq}_j ) \right),
    \end{eqnarray}
where $\alpha,\beta$ indicate the isotope type as above.

%%%%%%%%%%%%%%%%%%%%%%%%%%%%%%%%%%%%%%%%%%%%%%%%%%%%%%%%%%%%%%%%%%%
%%%%%%%%%%%%%%%%%%%%%%%%%%%%%%%%%%%%%%%%%%%%%%%%%%%%%%%%%%%%%%%%%%%

\subsection{The validity of the steady state}\label{sec:appendix:B}

\noindent
Equations \ref{eqn:multiocc-evolution} may be used to assess whether large changes in mobile concentration $x$ with time lead to large changes in the time derivative of the probability vector $\textbf{y}$. Taking the second derivative of equation \ref{eqn:time_derivative_y},

    \begin{eqnarray}
        \label{eqn:prob-second-deriv}                 
            \frac{\partial^2 {\bf y}}{\partial t^2} &=& - \frac{\partial}{\partial t}({\bf G}[x,T]) {\bf y} - \bf{G} \frac{\partial \bf{y}}{\partial t} \nonumber\\
             &=& - \frac{\partial \bf{G}}{\partial x}\frac{\partial x}{\partial t} \bf{y}+{\bf G}^2 \bf{y} \nonumber\\
             &=& \left[ {\bf G}^2 - \frac{\partial \bf{G}}{\partial x}\frac{\partial x}{\partial t} \right] {\bf{y}}.
    \end{eqnarray}

\noindent
We compute the spectral norm of the competing terms ${\bf G}^2$ and $\frac{\partial \bf{G}}{\partial x}\frac{\partial x}{\partial t}$ to estimate their maximum contribution in changing the time derivative of $\bf{y}$. For \textit{any} vector $\bf{y}$

\begin{equation}
    \left\|\frac{\partial \mathbf{G}}{\partial x}\frac{\partial x}{\partial t} \right\| \ll \|\mathbf{G}^2\|
\end{equation}
%\medskip

\noindent
must hold in order for the internal dynamics to dominate the evolution of $\bf{y}$, and not changes in the mobile concentration $x$ with time. The spectral norm of ${\bf G}^2$ is the square of the largest non-zero eigenvalue of ${\bf G}$. We recognise the smallest non-zero eigenvalue as the spectral gap $\mu(x,T)$ such that $\mu^2 \leq \|{\bf G}^2\|$. Therefore, a stricter condition on the time derivative of $x$ is

    \begin{equation}
        \frac{\partial x}{\partial t} \ll \frac{\mu(x,T)^2}{\left\|\frac{ \partial \bf{G}}{\partial x}\right\|} .
    \end{equation}

\noindent
Equation \ref{eqn:tridiagG} may be used to show that for a general $(n+1,n+1)$ $\bf{G}$ matrix, where $n$ is the maximum occupancy of the trap, the spectral norm of its derivative with respect to mobile fraction $x$ is bounded above by $2k$. We conclude

    \begin{equation}
    \label{eq:condition}
        \frac{\partial x}{\partial t} \ll \frac{\mu(x,T)^2}{2k}.
    \end{equation}

\noindent
For W at 600K and initial D mobile concentration $10^{-8}$ at. fr., the spectral gap according to figure \ref{fig:W_eigenvalues} leads to the condition $\frac{\partial x}{\partial t} \ll 10^{-8}$ at. fr. per second. For a flux $10^{18}$ m$^{-2}$s$^{-1}$, the source term would be $\sim 10^{-11}$ at. fr. during loading.

%%%%%%%%%%%%%%%%%%%%%%%%%%%%%%%%%%%%%%%%%%%%%%%%%%%%%%%%%%%%%%%%%%%
%%%%%%%%%%%%%%%%%%%%%%%%%%%%%%%%%%%%%%%%%%%%%%%%%%%%%%%%%%%%%%%%%%%

\subsection{A perturbation to the steady state}\label{sec:appendix:C}

\noindent
Suppose a trap begins far from steady state and condition \ref{eq:condition} holds for temperature $T$. The spectral gap $\mu$ gives the maximum time $\sim 1/\mu$ within which the steady state is reached. Assume the trap equilibrates with mobile concentration $x=x_0$. We can write the steady state eigenvalue $\lambda^{\rm eq} (x_0, T)=0$ and the left and right eigenvectors as $\textbf{z}^{\rm eq}\textbf{G}=\textbf{0}$ and $\textbf{G}\textbf{y}^{\rm eq}=\textbf{0}$ respectively. The structure of the general $\textbf{G}$ matrix in equation \ref{eqn:tridiagG} implies every element in the left eigenvector $\textbf{z}^{\rm eq}$ is one. \\

\noindent
We use the generalised Rayleigh quotient for non-Hermitian matrices, as formalised in \cite{Pattabhiraman1974}, to express the steady state eigenvalue as $\lambda ^{\rm eq} = \textbf{z}^{*,\rm eq}\textbf{G}\textbf{y}^{\rm eq}$, where $\textbf{z}^{*,\rm eq}$ is the conjugate transpose of the left eigenvector. With equation \ref{eqn:yeq_multiocc}, we note $\textbf{z}^{*,\rm eq} \textbf{y}^{\rm eq}=1$. To second order, the perturbation in eigenvalue can be written as:

\begin{equation}
    \lambda^{\rm eq}(x_0 +\Delta x) \approx \lambda^{\rm eq}(x_0) + \Delta x \left.\frac{d\lambda^{\rm eq}}{dx}\right|_{x_0} + \left.\frac{1}{2}\Delta x^2 \frac{d^2\lambda^{\rm eq}}{dx^2}\right|_{x_0}.
\end{equation}

\noindent
Applying the chain rule to the Rayleigh quotient, we write

\begin{align}
\label{eqn:first_RQ}
    \left.\frac{d\lambda^{\rm eq}}{dx} \right|_{x_0} &= \frac{d\textbf{z}^{*,\rm eq}}{dx}(\textbf{G}\textbf{y}^{\rm eq}) + \textbf{z}^{*,\rm eq} \frac{d\textbf{G}}{dx} \textbf{y}^{\rm eq} + (\textbf{z}^{*,\rm eq} \textbf{G})\frac{d\textbf{y}^{*,\rm eq}}{dx} \nonumber \\
    & = \textbf{z}^{*,\rm eq}(x_0) \frac{d\textbf{G}}{dx} \textbf{y}^{\rm eq}(x_0).
\end{align}

\noindent
With equations \ref{eqn:tridiagG} and \ref{eqn:yeq_multiocc}, it can be shown that $\left.{d\lambda^{\rm eq}}{dx} \right|_{x_0} = 0$. We may differentiate equation \ref{eqn:first_RQ} again with respect to $x$ to find the second-order perturbation $\left.{d^2\lambda^{\rm eq}}/{dx^2} \right|_{x_0}$. Given ${d\textbf{z}^{*,\rm eq}}/{dx}$ and ${d^2\textbf{G}}/{dx^2}$ are both zero, it is simple to show $\left.{d^2\lambda^{\rm eq}}/{dx^2} \right|_{x_0}=0$ also. From this, all higher derivatives are indeed also zero. Therefore, \textit{small} perturbations in mobile concentration $x$ do not push the system out of steady state. However, ${d\textbf{y}^{\rm eq}}/{dx} \neq \textbf{0}$ so the steady-state will indeed evolve with $x$.

%%%%%%%%%%%%%%%%%%%%%%%%%%%%%%%%%%%%%%%%%%%%%%%%%%%%%%%%%%%%%%%%%%%
%%%%%%%%%%%%%%%%%%%%%%%%%%%%%%%%%%%%%%%%%%%%%%%%%%%%%%%%%%%%%%%%%%%
%%%%%%%%%%%%%%%%%%%%%%%%%%%%%%%%%%%%%%%%%%%%%%%%%%%%%%%%%%%%%%%%%%%

\bibliographystyle{unsrt}
\bibliography{references}

\end{document}